\documentclass[journal]{IEEEtran}
\usepackage[hidelinks]{hyperref}
\usepackage{braket, physics, amssymb, amsthm}
\usepackage{bm}
\usepackage{slashed}
\usepackage[compat=1.1.0]{tikz-feynman}
\usepackage{tikz}
\usepackage{comment}
\usetikzlibrary{angles,quotes}
\usepackage{amsmath,amsthm,bm}

\usepackage[utf8]{inputenc}
\usepackage{multicol}
\usepackage{dcolumn}
\usepackage{verbatim}
\usepackage{soul}

\usepackage{xcolor}
\usepackage[normalem]{ulem}
\providecommand{\ignore}[1]{}

\newif\ifcmnt
\cmnttrue
\ifdefined\cmntsoff\cmntfalse\fi
\ifcmnt
    \providecommand{\aucmnt}[1]{#1}

\else
    \providecommand{\aucmnt}[1]{}

\fi

\usepackage[numbers,sort&compress]{natbib}
\begin{document}
\title{Uniform Additivity of Tripartite Optimized Correlation Measures}
\author{\IEEEauthorblockN{Joshua Levin\IEEEauthorrefmark{1}\IEEEauthorrefmark{4}, Ariel Shlosberg\IEEEauthorrefmark{1}\IEEEauthorrefmark{2}\IEEEauthorrefmark{4}, Vikesh Siddhu\IEEEauthorrefmark{1}, and Graeme Smith\IEEEauthorrefmark{1}\IEEEauthorrefmark{3}}\\ \hspace{10pt} \\
\small \emph{
\IEEEauthorblockA{\IEEEauthorrefmark{1}JILA,  University  of  Colorado/NIST,  Boulder, CO, 80309, USA
}\\
\IEEEauthorblockA{\IEEEauthorrefmark{2}Center for Quantum Information and Control (CQuIC), Department of Physics and Astronomy, University of
New Mexico, Albuquerque, New Mexico 87131, USA
}\\
\IEEEauthorblockA{\IEEEauthorrefmark{3}Institute for Quantum Computing and Department of Applied Mathematics, University of Waterloo, 200 University Ave W, Waterloo, ON N2L 3G1, Canada
}}
\IEEEcompsocitemizethanks{\IEEEcompsocthanksitem\IEEEauthorrefmark{4}Authors contributed equally}
}



\onecolumn
\maketitle
\begin{abstract}
    Information theory provides a framework for answering fundamental questions about the optimal performance of many important quantum communication and computational tasks. In many cases, the optimal rates of these tasks can be expressed in terms of regularized formulas that consist of linear combinations of von Neumann entropies optimized over state extensions. However, evaluation of regularized formulas is often intractable, since it involves computing a formula's value in the limit of infinitely many copies of a state. To find optimized, linear entropic functions of quantum states whose regularized versions are tractable to compute, we search for linear combinations of entropies on tripartite quantum states that are additive. We use the method of \cite{cross2017uniform}, which considers bipartite formulas, to identify convex polyhedral cones of uniformly additive \emph{tripartite} correlation measures. We rely only on strong subadditivity of the von Neumann entropy and use these cones to prove that three previously established tripartite optimized correlation measures are additive.
\end{abstract}

\IEEEpeerreviewmaketitle

\section{Introduction}
\IEEEPARstart{O}{ften}, fundamental bounds on performance in information-theoretic tasks can be expressed through linear combinations of subsystem entropies. As a simple example, the optimal rate at which a quantum state can be compressed is given by the state's von Neumann entropy, $S(\rho) = -\text{Tr}[\rho\log\rho]$ \cite{schumacher1995quantum}. For other types of tasks, the formulas describing optimal performance are more complicated and may involve optimizations over states of certain systems. For instance, the maximum rate at which classical information can be transmitted over a classical channel, which maps a message set $X$ to a random variable $Y$, is equal to the maximum value of the mutual information of $X$ and $Y$,
\begin{align}
    \max_{p(X)}I(X:Y) = \max_{p(X)}~[H(X) + H(Y) - H(XY)],\label{capacity}
\end{align}
where the message set $X$ is treated as a random variable with an associated distribution (or ``state") $p(X)$ that is optimized over in (\ref{capacity}) \cite{Shannon1948}. Here, $H(X) = -\sum_xp(x)\log p(x)$ is the Shannon entropy. Furthermore, for many tasks involving a fixed input state $\rho$, the formula expressing optimal performance involves optimizing a linear combination of entropies over the states of some auxiliary systems (referred to as ancillas), and the evaluation must be done in the limit of many copies of $\rho$. Consider the task of preparing a large number of copies, $n$, of a bipartite quantum state, $\rho_{AB}$, via LOCC operations on an initial finite supply of EPR pairs, subject to the restriction that the amount of communication between $A$ and $B$ is sublinear in $n$. One natural question to consider is what is the fewest number of EPR pairs per copy of $\rho_{AB}$ for which this task is possible? The answer is given by 
\begin{align}
E_P^\infty(\rho_{AB})\equiv\lim_{n\to\infty}\frac{1}{n}E_P(\rho_{AB}^{\otimes n})\label{reg EP},
\end{align}
where $E_P$ is the entanglement of purification \cite{terhal2002entanglement},
\begin{align}
    E_P(\rho_{AB}) = \min_{\rho_{ABV}}S(\rho_{AV}).
\end{align}
So-called regularized formulas like (\ref{reg EP}) are very challenging to compute in cases where the optimized formula being evaluated in the large $n$ limit is non-additive, which is likely the case for $E_P$ \cite{chen2012non}. For a function $E$ acting on a quantum state to be additive, we mean that
\begin{align}
    E(\rho_1\otimes\rho_2) = E(\rho_1) + E(\rho_2)
\end{align}
for any two states $\rho_1$ and $\rho_2$. When $E$ is additive, we have that $E^\infty \equiv \lim_{n\to\infty}\frac{1}{n}E(\rho^{\otimes n}) = E$, which means that computing the regularized version of $E$ is no more difficult than computing $E$ itself. In order to identify functions of states that have operational interpretations in terms of information theoretic tasks, and whose regularized evaluation might be tractable, it is useful to identify linear combinations of entropies that, when optimized over ancillas, yield additive functions. A systematic method for identifying such functions was introduced in \cite{cross2017uniform}, where the authors executed their method for functions of bipartite states\footnote{Actually, in \cite{cross2017uniform}, the authors prepare bipartite states by starting with a state on a single system $A$ and then acting on $\rho_A$ with a quantum channel $\mathcal{N}$, which produces a state $\rho_{BE}$ where $B$ is the channel's output and $E$ is the channel's environment.  The authors then search for functions $E$ of quantum channels for which $E(\mathcal{N}\otimes\mathcal{M}) = E(\mathcal{N}) + E(\mathcal{M})$.  This problem is equivalent to the problem of finding functions $E$ of bipartite states for which $E(\rho_1\otimes\rho_2) = E_P(\rho_1) + E_P(\rho_2)$.}. In this work, we will employ the same method for functions of tripartite states. In so doing, we are able to prove that three tripartite correlation measures introduced in \cite{dewolfe2020multipartite} -- $E_{c_2}$, $E_{b_1}$, and $E_{s_2}$ -- are additive\footnote{This naming scheme, although it may seem arcane, was established to reflect symmetries displayed by these quantities when evaluated using geometric methods in holographic states in AdS/CFT. See \cite{dewolfe2020multipartite} for details.}. Like in \cite{cross2017uniform}, all instances of additivity established in this work are proven using only strong subadditivity of the von Neumann entropy \cite{lieb1973proof}. 

Multipartite correlations in quantum states are much less understood than bipartite correlations, and have been the subject of a great deal of research \cite{walter2017multipartite}. While many open questions remain, it has been repeatedly demonstrated that multipartite entanglement plays an important role in a wide range of areas of study, including metrology~\cite{komar_quantum_2014,PhysRevA.97.042337}, many-body physics~\cite{RevModPhys.80.517,zeng2018quantum,doi:10.1126/science.1247715}, nonlocal games~\cite{10.1119/1.16503,PhysRevLett.106.020405}, quantum secret sharing \cite{Hillery_1999}, device-independent randomness certification \cite{PhysRevLett.132.080201}, and distributed quantum computing \cite{PhysRevA.59.4249,CALEFFI2024110672}. In the two-party case, questions about how to characterize entanglement and whether a system is maximally entangled have clear answers \cite{walter2017multipartite}. For bipartite pure states, there is only a single type of entanglement and the entanglement entropy, $S(\Tr_B[\ket{\psi_{AB}}\bra{\psi_{AB}}])$, quantifies the degree of entanglement present. However, already in the case of a three qubit system, the system can be entangled in two inequivalent ways \cite{D_r_2000}, and in the case of four qubits, there are already nine inequivalent entanglement classes \cite{Verstraete_2002}. Therefore, as a result of the richness and usefulness of multipartite entanglement, it is worthwhile to determine the set of tripartite correlation measures that have operational meanings. 

The remainder of this paper is organized as follows. In Sec. \ref{Prelims}, we review the necessary mathematical background, including optimized linear entropic formulas, and convex polyhedral cones. In Sec. \ref{Bipartite}, we review the methods and results of \cite{cross2017uniform}, which introduced the notion of \emph{uniform additivity} and identified the complete set of uniformly additive bipartite optimized linear entropic formulas. In Sec. \ref{Tripartite}, we apply the method of \cite{cross2017uniform} to identify uniformly additive \emph{tripartite} optimized linear entropic formulas. Finally, in Sec. \ref{Conclusion}, we end with some concluding remarks, and comments on future work.

\section{Preliminaries}\label{Prelims}
We will begin by reviewing some useful mathematical tools and definitions. First, we review the definition of \emph{optimized linear entropic formulas}. We then review the definition and properties of \emph{convex polyhedral cones} along with the related concept of the \emph{dual cone}.

\subsection{Optimized Linear Entropic Formulas}\label{OCMs}
A \emph{linear entropic formula} on parties $\{A_1,\dots,A_n\}$ is a function $f_\alpha$ that acts on an $n$-partite state $\rho_{A_1\cdots A_n}$ and takes the form
\begin{align}
    f_\alpha(\rho_{A_1\cdots A_n}) = \sum_{\emptyset\neq\mathcal{J}\subseteq\{A_1,\dots,A_n\}}\alpha_\mathcal{J}S(\rho_\mathcal{J}),
\end{align}
where $\alpha\in\mathbb{R}^{2^n-1}$, and $S$ is the von Neumann entropy\footnote{The components of $\alpha$ are arranged in \emph{shortlex} order, meaning that smaller subsets always precede larger subsets, and subsets of the same cardinality are ordered lexicographically.  For example, for $n=2$ we have $\alpha = (\alpha_A,\alpha_B,\alpha_{AB})$, and for $n=3$ we have $\alpha = (\alpha_A,\alpha_B,\alpha_C, \alpha_{AB},\alpha_{AC},\alpha_{BC},\alpha_{ABC})$.}.  Examples include the mutual information
\begin{align}
    I(A:B)\equiv f_{(1,1,-1)}(\rho_{AB}) = S(\rho_A) + S(\rho_B) - S(\rho_{AB})
\end{align}
for $n=2$, and the conditional mutual information
\begin{align}
    I(A:B|C)\equiv f_{(0,0,-1,0,1,1,-1)}(\rho_{ABC}) = S(\rho_{AC}) + S(\rho_{BC}) - S(\rho_{ABC}) - S(\rho_C)
\end{align}
for $n=3$.

An \emph{optimized linear entropic formula} on parties $\{A_1,\dots,A_n\}$ and ancilla systems $\{V_1,\cdots,V_m\}$ is a function $E_\alpha$ that acts on an $n$-partite state $\rho_{A_1\cdots A_n}$ and takes the form
\begin{align}
    E_\alpha(\rho_{A_1\cdots A_n}) = \min_{\rho_{A_1\dots A_nV_1\dots V_m}}f_\alpha(\rho_{A_1\dots A_nV_1\dots V_m}),
\end{align}
for some linear entropic formula $f_\alpha$ on $n+m$ parties.  The minimization is over all states $\rho_{A_1\dots A_nV_1\dots V_m}$ whose marginal on $A_1\dots A_n$ is equal to $\rho_{A_1\cdots A_n}$.  Such states are called \emph{extensions} of $\rho_{A_1\cdots A_n}$.  Examples for $n=2$ and $m=1$ include the entanglement of purification \cite{terhal2002entanglement},
\begin{align}
    E_P(A:B)\equiv E_{(0,0,0,0,1,0,0)}(\rho_{AB}) = \min_{\rho_{ABV}}S(\rho_{AV}),
\end{align}
and the squashed entanglement \cite{christandl2004squashed},
\begin{align}
    E_{sq}(A:B)\equiv E_{(0,0,-1,0,1,1,-1)}(\rho_{AB}) = \min_{\rho_{ABV}}I(A:B|V).
\end{align}
\subsection{Convex Polyhedral Cones}\label{Cones}
In this paper, the optimized linear entropic formulas that we are concerned with will form algebraic structures called convex polyhedral cones.  We will utilize the mathematical properties of convex polyhedral cones in order to identify additive optimized linear entropic formulas on three parties.
\subsubsection{Definitions and Properties}
Given a real vector space $V$, a \emph{linear cone} in $V$ is a subset $C\subseteq V$ with the property that $\alpha v\in C$ for all $v\in C$ and all $\alpha\in \mathbb{R}_+$.  One can then define a \emph{convex cone} as a linear cone with the additional property that $\alpha_1 v_1 + \alpha_2 v_2 \in C$ for all $v_1,v_2\in C$ and all $\alpha_1,\alpha_2\in \mathbb{R}_+$, \emph{i.e.} a linear cone which is also convex.  Given $v_1,v_2\in V$, the vector $\alpha_1 v_1 + \alpha_2 v_2$, for $\alpha_1,\alpha_2\in \mathbb{R}_+$ is called a \emph{conic combination} of $v_1$ and $v_2$, so one could equivalently define a convex cone as a subset of $V$ which is closed under conic combinations.  A convex cone $C$ is \emph{pointed} if $C$ contains no nontrivial linear subspace of $V$, or equivalently if $v,-v\in C$ implies $v=0$, or equivalently if $C\cap-C = \{0\}$.

Given a subset $S\subseteq V$, the \emph{conic hull} of $S$ is the set
\begin{align}
    \left\{\sum_{i=1}^k \alpha_i v_i ~\Bigg|~ \alpha_i \in \mathbb{R}_+,~ v_i \in S,~ k \in \mathbb{N}\right\}.
\end{align}
One can then define a \emph{polyhedral cone} as a convex cone $C$ which can be written as the conic hull of a finite subset $S\subset V$.  $S$ is called a generating set for $C$, and if no strict subset of $S$ is a generating set for $C$ then $S$ is called a \emph{minimal generating set}.  A polyhedral cone $C$ has a unique minimal generating set if and only if $C$ is pointed.  
\subsubsection{The Dual Cone} Any polyhedral cone $C$ has two equivalent descriptions.  One description is given above, where $C$ is characterized as the conic hull of a finite generating set $S$.  But $C$ also has finitely many planar faces $F_i$, each of which contains the origin and has an inward pointing normal vector $n_i$.  We can also characterize $C$ using this finite set $S^* = \{n_1,\dots,n_l\}$ of inward pointing normals, as follows.  Note that each face $F_i$, when extended to an infinite plane, divides the space $V$ into two halfspaces, one that we call $H_i^+$ which contains $C$, and one that we call $H_i^-$ whose intersection with $C$ is only $F_i$.  So for any $x\in C$ and any $n_i\in S^*$, we must have $x\in H_i^+$.  Therefore we can characterize $C$ as the intersection of all of the halfspaces $H_i^+$.  It is straightforward to see that $x\in H_i^+$ is equivalent to $x\cdot n_i\geq 0$, so we have
\begin{align}
    C = \{x~|~x\in H_i~~\forall~1\leq i\leq l\} = \{x~|~x\cdot n_i\geq 0~~\forall~1\leq i\leq l\}.\label{Cone}
\end{align}
We can think of the inequalities $x\cdot n_i\geq 0$ as a system of linear constraints on the components of $x$, which are all satisfied if and only if $x\in C$.  The dual cone $C^*$ can be defined as the polyhedral cone generated by $S^*$, or equivalently
\begin{align}
    C^* = \{x~|~x\cdot v\geq 0~~\forall~v\in C\}.\label{dual cone}
\end{align}
$C^*$ provides a dual description of $C$, which is equivalent to the minimal generating set description.  This dual description is given as a set of inequalities that are satisfied by the components of $x$ if and only if $x\in C$.  These inequalities are exactly the ones appearing ($\ref{dual cone}$), which are generated, via conic combinations, by the finite set of inequalities appearing in (\ref{Cone}).
\section{Uniform Additivity of Bipartite Optimized Linear Entropic Formulas}\label{Bipartite}
Let $E_\alpha$ be an optimized linear entropic formula which acts on a bipartite state $\rho_{AB}$, so 
\begin{align}
\alpha = (\alpha_A,\alpha_B,\alpha_V,\alpha_{AB},\alpha_{AV},\alpha_{BV},\alpha_{ABV}).
\end{align}
$E_\alpha$ is called \emph{additive} if
\begin{align}
    E_\alpha(\rho_{AB}) = E_\alpha(\rho_{A_1B_1}\otimes\rho_{A_2B_2}) = E_\alpha(\rho_{A_1B_1}) + E_\alpha(\rho_{A_2B_2})\label{additivity_2}
\end{align}
holds for any two states $\rho_{A_1B_1}$ and $\rho_{A_2B_2}$, where $A=A_1A_2$ and $B=B_1B_2$ in the product state in the middle expression\footnote{The way the middle expression is written, there is a potential confusion, which we hope this sentence clarifies. Whenever we refer to a vector $\alpha$, it is necessary to specify which subsystem is $A$ and which is $B$. In the expressions on the left and right of (\ref{additivity_2}), these subsystems are clear. But in the middle expression of (\ref{additivity_2}), we must specify that $A=A_1A_2$ and $B=B_1B_2$, lest the reader think that $A$ is the subsystem to the left of the $\otimes$ and $B$ is the subsystem to the right of the $\otimes$, in which case the value of $E_\alpha$ would be 0 in all cases.}.
To show that Eq. (\ref{additivity_2}) holds, we prove the two inequalities
\begin{align}
    E_\alpha(\rho_{A_1B_1}\otimes\rho_{A_2B_2}) &\leq E_\alpha(\rho_{A_1B_1}) + E_\alpha(\rho_{A_2B_2})\label{trivial_ineq}
    \\E_\alpha(\rho_{A_1B_1}\otimes\rho_{A_2B_2}) &\geq E_\alpha(\rho_{A_1B_1}) + E_\alpha(\rho_{A_2B_2}),\label{nontrivial_ineq}
\end{align}
known as subadditivity and superadditivity.  Let us first examine (\ref{trivial_ineq}).  Since each instance of $E_\alpha$ is a minimization of an objective function $f_\alpha$ over all state extensions, it suffices to choose two arbitrary extensions $\rho_{A_1B_1V_1}$ of $\rho_{A_1B_1}$ and $\rho_{A_2B_2V_2}$ of $\rho_{A_2B_2}$, and show that there exists an extension $\rho_{ABV}$ of $\rho_{AB}$ such that
\begin{align}
    f_\alpha(\rho_{ABV})\leq f_\alpha(\rho_{A_1B_1V_1}) + f_\alpha(\rho_{A_2B_2V_2}).\label{min1}
\end{align}
Given $\rho_{A_1B_1V_1}$ and $\rho_{A_2B_2V_2}$, let $\rho_{ABV} = \rho_{A_1B_1V_1}\otimes\rho_{A_2B_2V_2}$. Since $f_\alpha (\rho_{ABV})$ is a linear entropic formula evaluated on systems $A=A_1A_2$, $B=B_1B_2$, and $V=V_1V_2$, for any subset $\mathcal{I}\subseteq\{A,B,V\}$ we have
\begin{align}
    S(\rho_{\mathcal{I}}) = S(\rho_{\mathcal{I}_1}\otimes\rho_{\mathcal{I}_2}) = S(\rho_{\mathcal{I}_1}) + S(\rho_{\mathcal{I}_2}).
\end{align} 
Therefore (\ref{trivial_ineq}) is satisfied by all $\alpha$.
\bigskip
\subsection{Standard Decouplings and Uniform Additivity}\label{Standard decouplings}
To show that (\ref{nontrivial_ineq}) holds, it suffices to choose an arbitrary extension $\rho_{ABV}$ of $\rho_{AB}~(=\rho_{A_1B_1}\otimes\rho_{A_2B_2})$ and show that there exist extensions $\rho_{A_1B_1V_1}$ of $\rho_{A_1B_1}$ and $\rho_{A_2B_2V_2}$ of $\rho_{A_2B_2}$ such that
\begin{align}
    f_\alpha(\rho_{ABV})\geq f_\alpha(\rho_{A_1B_1V_1}) + f_\alpha(\rho_{A_2B_2V_2}).\label{min2}
\end{align}
A choice of $\rho_{A_1B_1V_1}$ and $\rho_{A_2B_2V_2}$, given $\rho_{ABV}$, is called a \emph{decoupling}. First note that since $\rho_{AB}=\rho_{A_1B_1}\otimes\rho_{A_2B_2}$, any entropy terms appearing in $f_\alpha$ that don't involve $V$ will have equal contributions on the left and right sides of (\ref{min2}), and are therefore not constrained by (\ref{min2}).  Any changes to the coefficients of these terms will preserve (non)additivity of $\alpha$. For this reason we only consider $\alpha$ with $\alpha_A = \alpha_B = \alpha_{AB} = 0$, so
\begin{align}
    \alpha = (\alpha_V, \alpha_{AV}, \alpha_{BV}, \alpha_{ABV}).
\end{align}

There is a constraint we can impose on $\alpha$ to guarantee the existence of a decoupling satisfying (\ref{min2}).  We can define $V_1 = \mathcal{S}_2V$ and $V_2 = \mathcal{S}_1V$ for subsets $\mathcal{S}_2\subseteq\{A_2,B_2\}$ and $\mathcal{S}_1\subseteq\{A_1,B_1\}$ so that
\begin{align}
    \rho_{A_1B_1V_1} &= \text{Tr}_{[A_2B_2\setminus\mathcal{S}_2]}(\rho_{ABV})
    \\\rho_{A_2B_2V_2} &= \text{Tr}_{[A_1B_1\setminus\mathcal{S}_1]}(\rho_{ABV}),
\end{align}
and require that (\ref{min2}), the resulting 5-party  entropy inequality on systems $\{A_1,B_1,A_2,B_2,V\}$, holds for all states $\rho_{A_1B_1A_2B_2V}$ which have the property that 
\begin{align}
    \text{Tr}_V(\rho_{A_1B_1A_2B_2V}) = \rho_{A_1B_1A_2B_2} = \rho_{A_1B_1}\otimes\rho_{A_2B_2}.\label{prod12}
\end{align}
The choice of subsets $(\mathcal{S}_2,\mathcal{S}_1)$ is called a \emph{standard decoupling}.  Any $\alpha$ that satisfies this constraint satisfies Eq. (\ref{additivity_2}) since it can be shown to satisfy (\ref{nontrivial_ineq}) using the $(\mathcal{S}_2,\mathcal{S}_1)$ standard decoupling.  Such an $E_\alpha$ is called \emph{uniformly additive}, since all entropies appearing in Ineqs. (\ref{min1}) and (\ref{min2}) are evaluated on marginals of a common state $\rho_{A_1B_1A_2B_2V}$.
The set of $\alpha$ which satisfy (\ref{min2}) via the $(\mathcal{S}_2,\mathcal{S}_1)$ decoupling for any $\rho_{ABV}$ satisfying (\ref{prod12}) can easily be seen to form a convex cone.  Suppose $\alpha_1$ and $\alpha_2$ are two such $\alpha$ and let $c_1,c_2\in\mathbb{R}_+$.  Then
\begin{align}
    f_{c_1\alpha_1 + c_2\alpha_2}(\rho_{ABV}) &=  c_1f_{\alpha_1}(\rho_{ABV}) + c_2f_{\alpha_2}(\rho_{ABV}) 
    \\&\geq c_1[f_{\alpha_1}(\rho_{A_1B_1V_1}) + f_{\alpha_1}(\rho_{A_2B_2V_2})] + c_2[f_{\alpha_2}(\rho_{A_1B_1V_1}) + f_{\alpha_2}(\rho_{A_2B_2V_2})]\label{cone}
    \\&= f_{c_1\alpha_1 + c_2\alpha_2}(\rho_{A_1B_1V_1}) + f_{c_1\alpha_1 + c_2\alpha_2}(\rho_{A_2B_2V_2}),
\end{align}
where we have used the positivity of $c_1$ and $c_2$ and superadditivity of $\alpha_1$ and $\alpha_2$ to write (\ref{cone}).  So we can associate a convex cone, denoted $\Pi^{(\mathcal{S}_2,\mathcal{S}_1)}$, of uniformly additive $\alpha$'s to each standard decoupling $(\mathcal{S}_2,\mathcal{S}_1)$.  Moreover, each $\alpha$ in $\Pi^{(\mathcal{S}_2,\mathcal{S}_1)}$ can be proven additive by using the $(\mathcal{S}_2,\mathcal{S}_1)$ decoupling to show that (\ref{min2}) holds for any $\rho_{ABV}$ satisfying (\ref{prod12}).

To simplify notation, index the subsets of $\{A,B\}$ according to the mapping
\begin{align}
    \{0,1,2,3\} \leftrightarrow \{\emptyset,A,B,AB\}.
\end{align}
For example, the $\{A_2,A_1B_1\}$ decoupling, wherein $V_1 = A_2V$ and $V_2 = A_1B_1V$, is now parameterized by the ordered pair $(1,3)$.  This is the same convention used in \cite{cross2017uniform}.

\subsection{Equivalence Classes of Standard Decouplings for 2 Parties and 1 Ancilla}\label{equivalence classes 2party}
Some pairs of standard decouplings will produce equivalent cones, either in the sense that the cones are exactly equal, or in the sense that they contain the same set of functions expressed as different $\alpha$ vectors.  The equivalence classes of uniform additivity cones are generated by three symmetries of the decouplings.  

First, we can swap the system labels 1 and 2.  This operation takes $\Pi^{(a,b)}$ to $\Pi^{(b,a)}$.  These two cones are exactly equal. To see this, recall that $\alpha$ in $\Pi^{(a,b)}$ means that
\begin{align}
    f_\alpha(\rho_{ABV})\geq f_\alpha(\rho_{A_1B_1(a_2V)}) + f_\alpha(\rho_{A_2B_2(b_1V)})\label{pi_ab}
\end{align}
for all states $\rho_{ABV}$ satisfying (\ref{prod12}).  So let $\alpha\in \Pi^{(a,b)}$ and choose any state $\rho_{ABV}$ which satisfies (\ref{prod12}).  Then let $\rho'_{ABV}$ be obtained from $\rho_{ABV}$ by swapping the system labels 1 and 2.  Since $\rho'_{ABV}$ also satisfies (\ref{prod12}), we must have
\begin{align}
    f_\alpha(\rho'_{ABV})\geq f_\alpha(\rho'_{A_2B_2(a_1V)}) + f_\alpha(\rho'_{A_1B_1(b_2V)}),\label{swap 12}
\end{align}
so $\alpha\in \Pi^{(b,a)}$.

Second, we can swap the system labels $A$ and $B$.\footnote{For optimized linear entropic formulas acting on $n$-partite states, this symmetry generalizes to any permutation of the $n$ systems.  We will see this in the $n=3$ case, which is addressed in the next section.}  This operation induces a permutation $P$ of the powerset of $\{A,B\}$, in particular
\begin{align}
   \{\emptyset,A,B,AB\}\to\{P(\emptyset),P(A),P(B),P(AB)\} = \{\emptyset,B,A,AB\},
\end{align}
as well as a permutation of the components of $\alpha$,
\begin{align}
    \alpha = (\alpha_V, \alpha_{AV}, \alpha_{BV}, \alpha_{ABV})\to P(\alpha) = (\alpha_V, \alpha_{BV}, \alpha_{AV}, \alpha_{ABV}),
\end{align}
which we also call $P$.  This operation takes $\Pi^{(a,b)}$ to $\Pi^{(P(a),P(b))}$.  These cones are not equal, but they contain the same set of functions up to a swap of the $A$ and $B$ arguments.  More precisely, for any $\alpha\in \Pi^{(a,b)}$ and $P(\alpha)\in \Pi^{(P(a),P(b))}$, we have 
\begin{align}
    E_\alpha(\rho_{AB}) = E_{P(\alpha)}(\rho_{BA})\label{swap AB}
\end{align}
for all states $\rho_{AB}$.

Third, we can include a purifying system $W$, and for each $\mathcal{I}\in\{V,AV,BV,ABV\}$, replace $\mathcal{I}$ with its complement $\Bar{\mathcal{I}}$ in $\{A,B,V,W\}$.  This operation takes $S(\rho_{AV})$ to $S(\rho_{BW})$, $S(\rho_{ABV})$ to $S(\rho_W)$, etc.  Note that since $\rho_{ABVW}$ is a pure state, this preserves the values of all entropies and therefore preserves the value of $E_\alpha$.  We can then simply rename $W$ as $V$. This operation induces a permutation $Q$ of the powerset of $\{A,B\}$,
\begin{align}
   \{\emptyset,A,B,AB\}\to\{Q(\emptyset),Q(A),Q(B),Q(AB)\} = \{AB,B,A,\emptyset\},
\end{align}
and a permutation of the components of $\alpha$
\begin{align}
    \alpha = (\alpha_V, \alpha_{AV}, \alpha_{BV}, \alpha_{ABV})\to Q(\alpha) = (\alpha_{ABV}, \alpha_{BV}, \alpha_{AV}, \alpha_V).
\end{align}
This operation takes $\Pi^{(a,b)}$ to $\Pi^{(Q(a),Q(b))} = \Pi^{(3-a,3-b)}$, which is not the same cone, but contains all the same functions since
\begin{align}
    E_\alpha(\rho_{AB}) = E_{Q(\alpha)}(\rho_{AB})\label{swap VW}
\end{align}
for all $\rho_{AB}$, again because $\rho_{ABVW}$ is pure.

These three symmetries generate five equivalence classes of standard decouplings:
\begin{enumerate}
    \item $\{(3,3),(0,0)\}$
    \item $\{(3,0),(0,3)\}$
    \item $\{(1,1),(2,2)\}$
    \item $\{(1,2),(2,1)\}$
    \item $\{(3,2),(2,3),(3,1),(1,3),(1,0),(0,1),(2,0),(0,2)\}$
\end{enumerate}
\bigskip
\subsection{The Additive Cones of Bipartite Optimized Linear Entropic Formulas}\label{Bipartite Cones}
To derive $\Pi^{(a,b)}$, we first generate a set of conditions on $\alpha$ which are necessary for $\alpha\in\Pi^{(a,b)}$.  Recall that if $\alpha\in\Pi^{(a,b)}$, then (\ref{pi_ab}) holds for all states $\rho_{ABV}$ satisfying (\ref{prod12}).  So if we choose any state $\rho_{ABV}$ that satisfies (\ref{prod12}) and evaluate all entropies appearing in (\ref{pi_ab}), then (\ref{pi_ab}) becomes a linear constraint on the components of $\alpha$, i.e. an inequality of the form
\begin{align}
    \alpha\cdot\beta(\rho)\geq 0,
\end{align}
which must be satisfied if $\alpha\in\Pi^{(a,b)}$. In particular, for $\rho = \rho_{ABV}$, defining
\begin{align}
    \Delta_{\mathcal{J},\rho}^{(a,b)} = S(\rho_{\mathcal{J}V}) - S(\rho_{\mathcal{J}_1V_1}) - S(\rho_{\mathcal{J}_2V_2})
\end{align}
for each $\mathcal{J}\subseteq\{A,B\}$, we have
\begin{align}
    \beta(\rho) = \left(\Delta_{\emptyset,\rho}^{(a,b)},~\Delta_{A,\rho}^{(a,b)},~\Delta_{B,\rho}^{(a,b)},~\Delta_{AB,\rho}^{(a,b)}\right).
\end{align} 
Each state $\rho_{ABV}$ satisfying (\ref{prod12}) will produce a necessary constraint $\beta$.  According to Lemma C.1 of \cite{cross2017uniform}, uniform additivity also requires $\alpha$ to be balanced in $V$.  More precisely, this means we must always include the constraints
\begin{align}
    \beta_+ &= (1,~1,~1,~1)
    \\\beta_- &= (-1,-1,-1,-1)
\end{align}
in our constraint set $B$, in addition to the necessary constraints generated using states.  Given $B = \{\beta_1,\dots, \beta_l\}$ generated by this method, we can then define the polyhedral cone $C(B)$ generated by $B$.  $C(B)$ is a cone of constraints on $\alpha$ which are necessary for $\alpha\in\Pi^{(a,b)}$.  The dual of $C(B)$, denoted $C^*(B)$, is the polyhedral cone consisting of all $\alpha$ which satisfy all constraints in $B$.  Therefore, we have 
\begin{align}
   \Pi^{(a,b)}\subseteq C^*(B).
\end{align}

To determine if the constraints $B$ are sufficient for
$\alpha\in\Pi^{(a,b)}$, i.e. whether
\begin{align}
    C^*(B)\subseteq\Pi^{(a,b)},
\end{align}
we consider a minimal generating set $A = \{\alpha_1,\dots,\alpha_k\}$ for $C^*(B)$.  We need only check that $\alpha_i\in\Pi^{(a,b)}$ for $1\leq i\leq k$.  To check that $\alpha_i\in\Pi^{(a,b)}$, we must verify that (\ref{pi_ab}) with $\alpha=\alpha_i$ (a 5-party entropy inequality on systems $\{A_1,B_1,A_2,B_2,V\}$) holds for all states $\rho_{ABV}$ satisfying (\ref{prod12}).  To do this, we use the known entropy inequalities from Appendix \ref{Entropy Ineqs}.  If this can be done for all $\alpha\in A$, then the constraints $B$ are not only necessary but also sufficient, and
\begin{align}
    \Pi^{(a,b)} = C^*(B).
\end{align}
If, for some $1\leq i\leq k$, (\ref{pi_ab}) with $\alpha=\alpha_i$ does not hold for some state $\rho_{ABV}$ satisfying (\ref{prod12}), then
\begin{align}
    \Pi^{(a,b)}\subset C^*(B),
\end{align} 
and the constraints $B$ are not sufficient for $\alpha\in\Pi^{(a,b)}$.  
Note that if $\Pi^{(a,b)}$ is not polyhedral, then no finite constraint set $B$ is sufficient, i.e. $\Pi^{(a,b)}\subset C^*(B)$ for any finite $B$.

This was done in \cite{cross2017uniform}\footnote{This was actually done in \cite{cross2017uniform} for \emph{maximized} linear entropic formulas, as opposed to \emph{minimized} linear entropic formulas, which we treat here.  The resulting cones differ by a factor of $-1$.}, and the resulting uniform additivity cones are indeed polyhedral and have the following minimal generating sets:
\begin{table}[h]
    \centering
    \begin{tabular}{c|c|c|c|c|c}
       Decoupling  & (3,3) & (3,0) & (1,1) & (1,2) & (3,2) \\
       \hline
       &&&&&\\
       Minimal&$S(A|BV)$&$-S(A|BV)$&$S(B|V)$&$-S(A|BV)$&$S(AB|V)$\\
       generating&$S(A|V)$&$S(A|V)$&$-S(A|BV)$&$S(A|V)$&$S(B|V)$\\
       set&$S(B|AV)$&$-S(AB|V)$& &$S(AV) - S(BV)$&$-S(B|AV)$\\
       &$S(B|V)$&$S(AB|V)$& &$S(BV) - S(AV)$&$S(B|AV)$\\
    \end{tabular}
    \vspace{10pt}
    \caption{Minimal generating sets for the uniform additivity cones of bipartite optimized linear entropic formulas with one ancilla.  Notice that the squashed entanglement is $(3,0)$-additive, $(1,2)$ additive, and $(3,2)$-additive.  Only one representative from each equivalence class of cones is shown, and the equivalent cones can be obtained by acting on the vectors shown with the appropriate permutation as described in Sec. \ref{equivalence classes 2party}.}
    \label{tab:2party_1_anc_cones}
\end{table}

Remarkably, for all of these cones, the sufficient constraint set $B$ can be generated using only classical states.  This means that the set of uniformly additive bipartite optimized linear entropic formulas acting on quantum states is identical to those acting on classical states!

\section{Uniform Additivity of Tripartite Optimized Linear Entropic Formulas}\label{Tripartite}
In this section, we will use the same method to identify the uniform additivity cones of tripartite optimized linear entropic formulas.  We consider formulas that are optimized over one ancilla $V$, so
\begin{align}
    \alpha = (\alpha_V,\alpha_{AV},\alpha_{BV},\alpha_{CV},\alpha_{ABV},\alpha_{ACV},\alpha_{BCV},\alpha_{ABCV}).
\end{align}
The resulting uniform additivity cones can be used to construct the uniform addivity cones for tripartite optimized linear entropic formulas optimized over any number of ancillas.
\subsection{Equivalence Classes of Standard Decouplings for 3 Parties and 1 Ancilla}\label{equivalence classes 3party 1ancilla}
We index the subsets of $\{A,B,C\}$ according to the mapping
\begin{align}
    \{0,1,2,3,4,5,6,7\}\leftrightarrow\{\emptyset,A,B,C,AB,AC,BC,ABC\},\label{subsets}
\end{align}
so the decouplings are now parameterized by ordered pairs $(a,b)$ with $0\leq a,b\leq7$.  
The symmetries of the associated uniform additivity cones are essentially the same as those of bipartite case, with two minor differences: (1) the symmetry generated by swapping $A$ and $B$ generalizes to a symmetry generated by acting on $\{A,B,C\}$ with any permutation, and (2) the symmetry generated by swapping $V$ and the purifying system $W$ now takes $\Pi^{(a,b)}$ to $\Pi^{(7-a,7-b)}$.  These symmetries generate the following eight equivalence classes of standard decouplings:
\begin{multicols}{4}
\begin{enumerate}
    \item $\begin{Bmatrix}
    (0,0)\\
    (7,7)
    \end{Bmatrix}$
    \newline
    \newline
    \newline
    \newline
    \newline
    \newline
    \item $\begin{Bmatrix}
    (0,7)\\
    (7,0)
    \end{Bmatrix}$
    \newline
    \newline
    \newline
    \newline
    \newline
    \item $\begin{Bmatrix}
    (1,1)\\
    (2,2)\\
    (3,3)\\
    (4,4)\\
    (5,5)\\
    (6,6)
    \end{Bmatrix}$
    \newline
    \newline
    \item $\begin{Bmatrix}
    (1,6)\\
    (2,5)\\
    (3,4)\\
    (4,3)\\
    (5,2)\\
    (6,1)
    \end{Bmatrix}$
    \newline
    \item $\begin{Bmatrix}
    (0,1) & (7,6)\\
    (0,2) & (7,5)\\
    (0,3) & (7,4)\\
    (1,0) & (6,7)\\
    (2,0) & (5,7)\\
    (3,0) & (4,7)
    \end{Bmatrix}$
    \newline
    \newline
    \item $\begin{Bmatrix}
    (0,4) & (7,3)\\
    (0,5) & (7,2)\\
    (0,6) & (7,1)\\
    (4,0) & (3,7)\\
    (5,0) & (2,7)\\
    (6,0) & (1,7)
    \end{Bmatrix}$
    \newline
    \item $\begin{Bmatrix}
    (1,2) & (6,5)\\
    (2,3) & (5,4)\\
    (3,1) & (4,6)\\
    (2,1) & (5,6)\\
    (3,2) & (4,5)\\
    (1,3) & (6,4)
    \end{Bmatrix}$
    \newline
    \newline
    \item $\begin{Bmatrix}
    (1,4) & (6,3)\\
    (2,6) & (5,4)\\
    (3,5) & (4,2)\\
    (2,4) & (5,3)\\
    (3,6) & (4,1)\\
    (1,5) & (6,2)
    \end{Bmatrix}$
\end{enumerate}
\end{multicols}
\subsection{The Additive Cones of Tripartite Optimized Linear Entropic Formulas with 1 Ancilla}\label{Tripartite 1ancilla cones}
In the tripartite case, any state $\rho_{ABCV}$ which satisfies
\begin{align}
    \text{Tr}_V(\rho_{ABCV}) = \rho_{A_1B_1C_1A_2B_2C_2} = \rho_{A_1B_1C_1}\otimes\rho_{A_2B_2C_2}\label{tri_prod12},
\end{align}
generates a necessary condition for $\alpha\in\Pi^{(a,b)}$.  Motivated by the result from \cite{cross2017uniform} that any bipartite optimized linear entropic formula which is uniformly additive on all classical states is also uniformly additive on all quantum states, we chose to generate our necessary constraint set using a particular set of classical states.  The 7-party states in this set have three properties:
\begin{enumerate}
    \item All 7 parties are uniform 
    classical bits or trivial 0-entropy random variables, or all 7 parties are uniform classical trits or trivial 0-entropy random variables.
    \item Within systems 1 and 2, any subset of the 3 parties can be maximally correlated with any disjoint subset.  For example, in a 7-bit state $\rho_{ABCV}$, we can have
    \begin{align}
    A_1 = B_1~~~~~\text{and}~~~~~B_2 = A_2\oplus_2 C_2,
    \end{align}
    or in a 7-trit state we can have
    \begin{align}
        B_2 = A_2\oplus_3 2C_2~~~~~\text{and}~~~~~A_1 = B_1 = C_1.
    \end{align}
    But since all states must satisfy Eq. (\ref{tri_prod12}), we do not allow any correlations between systems 1 and 2, such as
    \begin{align}
        A_1 = B_2 \oplus_2 C_2.
    \end{align}
    \item The ancilla system $V$ is allowed to be maximally correlated with any subset of the other six parties.  For example, we can have
    \begin{align}
        V = A_1 \oplus_2 B_1 \oplus_2 A_2 \oplus_2 B_2 
    \end{align}
    for a 7-bit state, or
    \begin{align}
        V = A_1 \oplus_3 A_2 \oplus_3 B_2 \oplus_3 2C_2
    \end{align}
    for a 7-trit state.
\end{enumerate}
For five of the eight equivalence classes, the resulting constraint set $B$ is sufficient for $\alpha\in\Pi^{(a,b)}$, so
\begin{align}
    \Pi^{(a,b)} = C^*(B)
\end{align}
for those cases.  In the other three cases, $C^*(B)$ contains at least one $\alpha\notin\Pi^{(a,b)}$.  For these cases, $C^*(B)$ is an outer bound for $\Pi^{(a,b)}$, and $C(A)$ is an inner bound if $A$ is any set of vectors known to be in $\Pi^{(a,b)}$ by satisfying 
\begin{align}
    f_\alpha(\rho_{ABCV})\geq f_\alpha(\rho_{A_1B_1C_1(a_2V)}) + f_\alpha(\rho_{A_2B_2C_2(b_1V)})\label{tri_pi_ab}
\end{align}
for all states $\rho_{ABCV}$ which satisfy (\ref{tri_prod12}).

We now choose one representative from each equivalence class, and show the uniform additivity cones that result from executing the method described above. These results are summarized in Tables \ref{tab:1_anc_cones} and \ref{tab:1_anc_cones_inout}. For each decoupling $(a,b)$, we move all terms in (\ref{tri_pi_ab}) to the right side and denote the resulting quantity $\Delta^{V,(a,b)}$ (consistent with the notation of \cite{cross2017uniform}) so that (\ref{tri_pi_ab}) becomes $\Delta^{V,(a,b)}\leq 0$. $\mathcal{B}$ represents a uniform bit and $\mathcal{T}$ represents a uniform trit. As a representative example, we show the construction of the additivity cone associated to the $(0,0)$ decoupling. The following relation defines the cone:
\begin{gather}
\label{eq:00reference}
\hspace*{-35pt}
        \Delta^{V,(0,0)} = \alpha_{AV} I(A_1:A_2|V)
        +\alpha_{BV} I(B_1:B_2|V)
        +\alpha_{CV} I(C_1:C_2|V)
        +\alpha_{ABV} I(A_1B_1:A_2B_2|V)\\
        +\alpha_{ACV} I(A_1C_1:A_2C_2|V)
        +\alpha_{BCV} I(B_1C_1:B_2C_2|V)
        +\alpha_{ABCV} I(A_1B_1C_1:A_2B_2C_2|V)\leq 0.
        \nonumber
    \end{gather}
    We now need necessary and sufficient conditions on $\alpha$ to satisfy the inequality given by Eq. \ref{eq:00reference}.
    \subitem \textit{Necessary conditions:}  The conditions
    \begin{subequations}
    \begin{align}
        \alpha_{ABCV} + \alpha_{ABV} + \alpha_{ACV} + \alpha_{BCV} + \alpha_{AV} + \alpha_{BV} + \alpha_{CV} &\leq 0 \label{1}\\
        \alpha_{ABCV} + \alpha_{ABV} + \alpha_{ACV} + \alpha_{BCV} + \alpha_{AV} + \alpha_{BV} &\leq 0 \label{16}\\
        \alpha_{ABCV} + \alpha_{ABV} + \alpha_{ACV} + \alpha_{BCV} + \alpha_{AV} + \alpha_{CV} &\leq 0 \label{17}\\
        \alpha_{ABCV} + \alpha_{ABV} + \alpha_{ACV} + \alpha_{BCV} + \alpha_{BV} + \alpha_{CV} &\leq 0 \label{18}\\
        \alpha_{ABCV} + \alpha_{ABV} + \alpha_{ACV} + \alpha_{BCV} + \alpha_{AV} &\leq 0 \label{13}\\
        \alpha_{ABCV} + \alpha_{ABV} + \alpha_{ACV} + \alpha_{BCV} + \alpha_{BV} &\leq 0 \label{14}\\
        \alpha_{ABCV} + \alpha_{ABV} + \alpha_{ACV} + \alpha_{BCV} + \alpha_{CV} &\leq 0 \label{15}\\
        \alpha_{ABCV} + \alpha_{ABV} + \alpha_{ACV} + \alpha_{BCV} &\leq 0\label{12}\\
        \alpha_{ABCV} + \alpha_{ABV} + \alpha_{ACV} + \alpha_{AV} &\leq 0\label{2}\\
        \alpha_{ABCV} + \alpha_{ABV} + \alpha_{BCV} + \alpha_{BV} &\leq 0\label{3}\\
        \alpha_{ABCV} + \alpha_{ACV} + \alpha_{BCV} + \alpha_{CV} &\leq 0\label{4}\\
        \alpha_{ABCV} + \alpha_{ACV} + \alpha_{BCV} &\leq 0\label{9}\\
        \alpha_{ABCV} + \alpha_{ABV} + \alpha_{BCV} &\leq 0\label{10}\\
        \alpha_{ABCV} + \alpha_{ABV} + \alpha_{ACV} &\leq 0\label{11}\\
        \alpha_{ABCV} + \alpha_{ABV} &\leq 0\label{5}\\
        \alpha_{ABCV} + \alpha_{ACV} &\leq 0\label{6}\\
        \alpha_{ABCV} + \alpha_{BCV} &\leq 0\label{7}\\
        \alpha_{ABCV} &\leq 0\label{8}
    \end{align}\label{00_ineq}
    \end{subequations}
    are necessary for non-positivity of $\Delta^{V,(0,0)}$. These inequalities are implied by the following classical states:
    \begin{subequations}
    \begin{align}
    &A_1 = B_1 = C_1 = \mathcal{B}_1&&A_2 = B_2 = C_2 = \mathcal{B}_2&&&V = \mathcal{B}_1\oplus \mathcal{B}_2
    \\&A_1 = B_1 = \mathcal{B}_1&&A_2 = B_2 = \mathcal{B}_2&&&V = \mathcal{B}_1\oplus \mathcal{B}_2
    \\&A_1 = C_1 = \mathcal{B}_1&&A_2 = C_2 = \mathcal{B}_2&&&V = \mathcal{B}_1\oplus \mathcal{B}_2
    \\&B_1 = C_1 = \mathcal{B}_1&&B_2 = C_2 = \mathcal{B}_2&&&V = \mathcal{B}_1\oplus \mathcal{B}_2
    \\&A_1 = \mathcal{B}_1, B_1 = \mathcal{B}_2,C_1 = A_1\oplus B_1&&A_2 = \mathcal{B}_3, B_2 = \mathcal{B}_4,C_2 = A_2\oplus B_2&&&V = \mathcal{B}_1\oplus \mathcal{B}_3
    \\&B_1 = \mathcal{B}_1, C_1 = \mathcal{B}_2,A_1 = B_1\oplus C_1&&B_2 = \mathcal{B}_3, C_2 = \mathcal{B}_4,A_2 = B_2\oplus C_2&&&V = \mathcal{B}_1\oplus \mathcal{B}_3
    \\&C_1 = \mathcal{B}_1, A_1 = \mathcal{B}_2,B_1 = C_1\oplus A_1&&C_2 = \mathcal{B}_3, A_2 = \mathcal{B}_4,B_2 = C_2\oplus A_2&&&V = \mathcal{B}_1\oplus \mathcal{B}_3
    \\&A_1 = \mathcal{T}_1, B_1 = \mathcal{T}_2,C_1 = A_1\oplus B_1&&A_2 = \mathcal{T}_3, B_2 = \mathcal{T}_4,C_2 = A_2\oplus B_2&&&V = \mathcal{T}_1\oplus\mathcal{T}_3\oplus -\mathcal{T}_2\oplus -\mathcal{T}_4
    \\&A_1 = \mathcal{B}_1, B_1 = \mathcal{B}_2, C_1 = \mathcal{B}_3&&A_2 = \mathcal{B}_4, B_2 = \mathcal{B}_5, C_2 = \mathcal{B}_6&&&V = \mathcal{B}_1\oplus\mathcal{B}_4
    \\&A_1 = \mathcal{B}_1, B_1 = \mathcal{B}_2, C_1 = \mathcal{B}_3&&A_2 = \mathcal{B}_4, B_2 = \mathcal{B}_5, C_2 = \mathcal{B}_6&&&V = \mathcal{B}_2\oplus\mathcal{B}_5
    \\&A_1 = \mathcal{B}_1, B_1 = \mathcal{B}_2, C_1 = \mathcal{B}_3&&A_2 = \mathcal{B}_4, B_2 = \mathcal{B}_5, C_2 = \mathcal{B}_6&&&V = \mathcal{B}_3\oplus\mathcal{B}_6
    \\&A_1 = B_1 = \mathcal{B}_1, C_1 = \mathcal{B}_2&&A_2 = B_2 = C_2 = \mathcal{B}_3&&&V = \mathcal{B}_1\oplus\mathcal{B}_2\oplus\mathcal{B}_3
    \\&A_1 = C_1 = \mathcal{B}_1, B_1 = \mathcal{B}_2&&A_2 = B_2 = C_2 = \mathcal{B}_3&&&V = \mathcal{B}_1\oplus\mathcal{B}_2\oplus\mathcal{B}_3
    \\&B_1 = C_1 = \mathcal{B}_1, A_1 = \mathcal{B}_2&&A_2 = B_2 = C_2 = \mathcal{B}_3&&&V = \mathcal{B}_1\oplus\mathcal{B}_2\oplus\mathcal{B}_3
    \\&A_1 = \mathcal{B}_1, B_1 = \mathcal{B}_2, C_1 = \mathcal{B}_3&&A_2 = \mathcal{B}_4, B_2 = \mathcal{B}_5, C_2 = \mathcal{B}_6&&&V = \mathcal{B}_1\oplus\mathcal{B}_2\oplus\mathcal{B}_4\oplus\mathcal{B}_5
    \\&A_1 = \mathcal{B}_1, B_1 = \mathcal{B}_2, C_1 = \mathcal{B}_3&&A_2 = \mathcal{B}_4, B_2 = \mathcal{B}_5, C_2 = \mathcal{B}_6&&&V = \mathcal{B}_1\oplus\mathcal{B}_3\oplus\mathcal{B}_4\oplus\mathcal{B}_6
    \\&A_1 = \mathcal{B}_1, B_1 = \mathcal{B}_2, C_1 = \mathcal{B}_3&&A_2 = \mathcal{B}_4, B_2 = \mathcal{B}_5, C_2 = \mathcal{B}_6&&&V = \mathcal{B}_2\oplus\mathcal{B}_3\oplus\mathcal{B}_5\oplus\mathcal{B}_6
    \\&A_1 = \mathcal{B}_1, B_1 = \mathcal{B}_2, C_1 = \mathcal{B}_3&&A_2 = \mathcal{B}_4, B_2 = \mathcal{B}_5, C_2 = \mathcal{B}_6&&&V = \bigoplus\mathcal{B}_i
    \end{align}
    \end{subequations}
    \subitem \textit{Sufficient conditions:}  These conditions are also sufficient.
    
    Applying the balancing condition and dualizing (\ref{00_ineq}), we find the minimal generating set of the (0,0) cone:
     \begin{align}
        (0,0,0,0,0,0,1,-1) &\to -S(A|BCV) \nonumber\\
        (0,0,0,0,0,1,0,-1) &\to -S(B|ACV) \nonumber\\
        (0,0,0,0,1,0,0,-1) &\to -S(C|ABV) \nonumber\\
        (0,0,1,0,-1,0,0,0) &\to -S(A|BV) \nonumber\\
        (0,0,0,1,0,-1,0,0) &\to -S(A|CV) \nonumber\\
        (0,1,0,0,-1,0,0,0) &\to -S(B|AV) \nonumber\\
        (0,0,0,1,0,0,-1,0) &\to -S(B|CV) \nonumber\\
        (0,1,0,0,0,-1,0,0) &\to -S(C|AV) \nonumber\\
        (0,0,1,0,0,0,-1,0) &\to -S(C|BV) \nonumber\\
        (1,-1,0,0,0,0,0,0) &\to -S(A|V) \nonumber\\
        (1,0,-1,0,0,0,0,0) &\to -S(B|V) \nonumber\\
        (1,0,0,-1,0,0,0,0) &\to -S(C|V) \nonumber
    \end{align}

Table \ref{tab:1_anc_cones} shows one representative $(a,b)$ from each of the five equivalence classes we were able to fully solve, and a minimal generating set for $\Pi^{(a,b)}$.  Table \ref{tab:1_anc_cones_inout} shows one representative $(a,b)$ from each of the three equivalence classes to which our method did not find an exact solution, and minimal generating sets for the best known inner and outer bounds for $\Pi^{(a,b)}$.

\begin{table}[h]
    \centering
    \begin{tabular}{c|c|c|c|c|c}
       Decoupling  & (0,0) & (7,0) & (1,6) & (1,0) & (4,0) \\
       \hline
       &&&&&\\
       &$-S(A|BCV)$&$S(AB|V)$&$-S(A|BCV)$&$-S(A|BCV)$&$-S(ABC|V)$\\
       &$-S(B|ACV)$&$S(AC|V)$&$-S(B|ACV)$&$-S(B|ACV)$&$-S(AB|CV)$\\
       Minimal&$-S(C|ABV)$&$S(BC|V)$&$-S(C|ABV)$&$-S(C|ABV)$&$-S(A|BCV)$\\
       generating&$-S(A|BV)$&$S(A|V)$&$S(BC|V)$&$-S(AC|V)$&$-S(B|ACV)$\\
       set&$-S(A|CV)$&$S(B|V)$&$S(B|V)$&$-S(AB|V)$&$S(A|V)$\\
       &$-S(B|AV)$&$S(C|V)$&$S(C|V)$&$-S(A|CV)$&$S(B|V)$\\
       &$-S(B|CV)$&$-S(ABC|V)$&$S(AV) - S(BCV)$&$-S(A|BV)$&$S(AB|V)$\\
       &$-S(C|AV)$&$S(ABC|V)$&$S(BCV) - S(AV)$&$-S(A|V)$&$-S(AB|V)$\\
       &$-S(C|BV)$&&&$S(A|V)$&\\
       &$-S(A|V)$&&&&\\
       &$-S(B|V)$&&&&\\
       &$-S(C|V)$&&&&\\
    \end{tabular}
    \vspace{10pt}
    \caption{Minimal generating sets for the uniform additivity cones of five inequivalent standard decouplings.  Again, only one representative from each equivalence class of cones is shown, and the equivalent cones can be obtained by acting on the vectors shown with the appropriate permutation as described in Sec. \ref{equivalence classes 3party 1ancilla}.}
    \label{tab:1_anc_cones}
\end{table}

\begin{table}[h]
    \centering
    \begin{tabular}{c|c|c|c}
        Decoupling & (1,1) & (1,2) & (1,4)\\
        \hline
        &&&\\
        Minimal&$S(A|V)$&$S(A|V)$&$S(A|V)$\\
        generating set&$-S(B|AV)$&$-S(B|AV)$&$-S(B|AV)$\\
        for inner&$-S(C|AV)$&$-S(A|BCV)$&$S(B|AV)$\\
        bound&$-S(B|ACV)$&$-S(B|ACV)$&$-S(B|ACV)$\\
        &$-S(C|ABV)$&$-S(C|ABV)$&$-S(A|BCV)$\\
        &&$S(AV) - S(BV)$&$-S(BC|AV)$\\
        &&$S(BV) - S(AV)$&\\
        &&&\\
        \hline
        &&&\\
        Additional&$I(A:B|V)$&$I(A:C|V) - S(B|ACV)$&$I(A:C|V) - S(B|ACV)$\\
        vectors for&$I(A:C|V)$&&$I(A:BC|V)$\\
        outer bound&$I(A:BC|V)$&&
    \end{tabular}
    \vspace{10pt}
    \caption{Minimal generating sets for the best known inner and outer bounds for the uniform additivity cones of the three remaining inequivalent standard decouplings.  The first row gives the minimal generating sets for the inner bounds, and appending the vectors in the second row gives the minimal generating sets for the outer bounds.}
    \label{tab:1_anc_cones_inout}
\end{table}

For the cones that were not fully solvable, we found a set of necessary inequalities whose dual is a supercone of the uniformly additive cone. This supercone is a strict outer bound as it contains extreme rays that provably violate SSA. We are able to prove the violations by having PSITIP \cite{li2021automated} generate counterexamples to the associated inequalities. However, the inequalities generated by these counterexamples involve irrational coefficients, leading to numerical difficulties in computing the dual cone, and therefore in tightening these bounds.


\subsection{Optimized Tripartite Linear Entropic Formulas with 2 Ancillas}\label{Tripartite OCMs}
Tripartite linear entropic formulas which are optimized over 2 ancilla systems $V$ and $W$ can contain 24 different entropy terms\footnote{There are also 7 entropies which do not contain either of the ancilla systems, but like in the previous cases these are not constrained by requiring additivity, so we set the coefficients of these entropies to 0 in our analysis.}: 8 involving only the ancilla system $V$ and some subset of $\{A,B,C\}$, 8 involving only the ancilla system $W$ and some subset of $\{A,B,C\}$, and 8 involving both $V$ and $W$ and some subset of $\{A,B,C\}$.  We can use this structure to break up this 24-dimensional vector into a direct sum of three vectors, $\alpha = \alpha^V\oplus\alpha^W\oplus\alpha^{VW}$, where
\begin{align}
    \alpha^V &= (\alpha_V,\alpha_{AV},\alpha_{BV},\alpha_{CV},\alpha_{ABV},\alpha_{ACV},\alpha_{BCV},\alpha_{ABCV})\\
    \alpha^W &= (\alpha_W,\alpha_{AW},\alpha_{BW},\alpha_{CW},\alpha_{ABW},\alpha_{ACW},\alpha_{BCW},\alpha_{ABCW})\\
    \alpha^{VW} &= (\alpha_{VW},\alpha_{AVW},\alpha_{BVW},\alpha_{CVW},\alpha_{ABVW},\alpha_{ACVW},\alpha_{BCVW},\alpha_{ABCVW}).
\end{align}
The decouplings that prove additivity for such formulas are parameterized by a pair of ordered pairs $(a,b)(c,d)$.  The symbols $\{a,b,c,d\}$ still take integer values from 0 to 7 and refer to subsets of $\{A,B,C\}$ via (\ref{subsets}).  Symbols $a$ and $b$ refer respectively to subsets of $\{A_2,B_2,C_2\}$ and $\{A_1,B_1,C_1\}$, which when combined with $V$ form the systems $V_1$ and $V_2$. Similarly, $c$ and $d$ refer respectively to subsets of $\{A_2,B_2,C_2\}$ and $\{A_1,B_1,C_1\}$, which when combined with $W$ form the systems $W_1$ and $W_2$.  We require these decouplings to be \emph{consistent}, meaning that $a$ and $c$ must be disjoint, and $b$ and $d$ must be disjoint.  This is to ensure that the two instances of $f_\alpha$ on the right side of the tripartite 2-ancilla version of (\ref{pi_ab}),
\begin{align}
    f_\alpha(\rho_{ABCVW})\geq f_\alpha(\rho_{A_1B_1C_1(a_2V)(c_2W)}) + f_\alpha(\rho_{A_2B_2C_2(b_1V)(d_1W)})\label{pi_abcd}
\end{align}
are evaluated on marginals of the larger state $\rho_{ABCVW}$, and (\ref{pi_abcd}) therefore forms a genuine 8-party entropy inequality (on parties $\{A_1,B_1,C_1,A_2,B_2,C_2,V,W\}$). 
\subsection{Equivalence Classes of Consistent Standard Decouplings for 3 Parties and 2 Ancillas}\label{Eqivalence classes 3party 2ancilla}
Like in the previous cases, the consistent standard decouplings for 3 parties and 2 ancillas have three symmetries which separate them into equivalence classes.  The first two symmetries are the same as in the 3 parties and 1 ancilla case: swapping systems 1 and 2 takes $\Pi^{(a,b)(c,d)}$ to $\Pi^{(b,a)(d,c)}$, which is exactly the same cone by an argument analogous to (\ref{swap 12}), and permuting $\{A,B,C\}$, which induces a permutation $P$ of the powerset of $\{A,B,C\}$, takes $\Pi^{(a,b)(c,d)}$ to $\Pi^{(P(a),P(b))(P(c),P(d))}$, which is an equivalent cone in a way analagous to (\ref{swap AB}).  The only difference is that the symmetry operation which swapped $V$ and the purifying system $W$ in the previous two cases now generalizes to include any permutation of the ancillas $V$ and $W$ and the purifying system $X$.  This operation induces a permutation $Q$ on the powerset of $\{A,B,C\}$, and takes $\Pi^{(a,b)(c,d)}$ to $\Pi^{(Q(a),Q(b))(Q(c),Q(d))}$ which is an equivalent cone, in a way analogous to (\ref{swap VW}).  These symmetries generate 22 equivalence classes of consistent standard decouplings, which are listed in the Appendix.
\subsection{Additive cones of Tripartite Formulas with 2 Ancillas from the Additive cones of Tripartite Formulas with 1 Ancilla}\label{3party 2ancilla cones}
According to Theorem G.1 of \cite{cross2017uniform}, $\alpha\in\Pi^{(a,b),(c,d)}$ if and only if $\alpha^V\in\Pi^{(a,b)}$, $\alpha^W\in\Pi^{(c,d)}$, and $\alpha^{VW}\in\Pi^{(a\cup c,b\cup d)}$.  Therefore, any 3 party and 2 ancilla uniform additivity cone $\Pi^{(a,b)(c,d)}$ can be constructed as the direct sum of three 1 ancilla cones,
\begin{align}
    \Pi^{(a,b)(c,d)} = \Pi^{(a,b)}\oplus\Pi^{(c,d)}\oplus\Pi^{(a\cup c,b\cup d)}.\label{direct sum}
\end{align}
As an application of (\ref{direct sum}), we will now determine which of the tripartite optimized linear entropic formulas introduced in \cite{dewolfe2020multipartite} (objective function $f_\alpha$ shown in the third column of Table \ref{tripartite OCMs} in the Appendix) are uniformly additive.  First note that these tripartite formulas are optimized over three ancilla systems denoted $a$, $b$, and $c$, but the full extension $\rho_{ABCabc}$ is required to be pure.  This is equivalent to optimizing over only two ancillas $V$ and $W$, and not requiring the extension $\rho_{ABCVW}$ to be pure.  This is because any optimized linear entropic formula which is optimized over pure extensions can be written in a way that does not depend on one of the ancillas, say $c$, by simply replacing any entropy involving $c$ with the entropy of its complement.  So for each optimized linear entropic formula $E_\alpha$ in the table, there are 6 ways to rewrite $E_\alpha$ using only two ancillas.  There are 3 options for which ancilla to eliminate, and 2 options for how to assign labels $V$ and $W$ to the two remaining ancillas.  From each 2-ancilla formula we can then extract the 8-dimensional vectors $\alpha^V$, $\alpha^W$, and $\alpha^{VW}$.  We can then use these vectors to determine, via (\ref{direct sum}), which of the uniform additivity cones $\Pi^{(a,b)(c,d)}$ contain a 2-ancilla version of $E_\alpha$. The results of this procedure are that $E_{c_2}$, $E_{b_1}$, and $E_{s_2}$ are uniformly additive. All others in Table \ref{tripartite OCMs} either lie outside of all of the exactly solved uniform additivity cones and outside of the known outer bounds of the remaining uniform additivity cones, or lie between the inner and outer bounds of a uniform additivity cone $\Pi^{(a,b)(c,d)}$ but are not $(a,b)(c,d)$-additive because the $(a,b)(c,d)$ decoupling does not produce a valid proof of additivity.

\section{Closing Remarks}\label{Conclusion}
We have extended the formalism introduced in \cite{cross2017uniform} for the bipartite case, to finding convex polyhedral cones containing tripartite correlation measures that are uniformly additive when optimized over extensions. This procedure allowed us to prove that three of the tripartite correlation measures introduced in \cite{dewolfe2020multipartite} -- $E_{c_2}$, $E_{b_1}$, and $E_{s_2}$ -- are uniformly additive. This implies that their regularized versions are tractable to compute and may yield operational interpretations for interesting quantum tasks involving three parties. We have established that the rest of the tripartite correlation measures in Table \ref{tripartite OCMs} are not uniformly additive, suggesting that these measures are non-additive. 

We were able to exactly solve five of the eight uniform additivity cones and establish inner and outer bounds for the remaining three cones. We leave the problem of tightening the bounds for these three cones to future work. 
For the exactly-solved cones, just as in the bipartite case, the set of tripartite quantities that are uniformly additive on quantum states is the same as the set of quantities that are uniformly additive on classical states. Exactly solving the remaining three cones could be possible by only considering more complex classical states, but may require considering constraints generated from quantum states. If it turns out to be the case that one needs quantum states to exactly solve for the uniform additivity cones, then this would signal a very significant difference in the role of quantum distributions between the tripartite and bipartite cases. 

In principle, the formalism used in this work can also be extended to consider the additivity of multipartite correlation measures for more than three parties. However, the computation of dual cones presents a significant computational bottleneck, since the dimension of the uniform additivity cones grows exponentially in the number of parties.

\section{Acknowledgements}
JL thanks Mohammad Alhejji for helpful discussions.  Throughout this work, we use PSITIP (introduced by Cheuk Ting Li in \cite{li2021automated}) to verify or disprove various entropy inequalities. This work was funded by NSF Grant No. 1734006 and NSF CAREER Grant No. CCF 1652560.
\bibliography{Uni_Add_TriOCM}

\begin{thebibliography}{10}

\bibitem{cross2017uniform}
A.~Cross, K.~Li, and G.~Smith, ``Uniform additivity in classical and quantum
  information,'' {\em Physical review letters}, vol.~118, no.~4, p.~040501,
  2017.

\bibitem{schumacher1995quantum}
B.~Schumacher, ``Quantum coding,'' {\em Physical Review A}, vol.~51, no.~4,
  p.~2738, 1995.

\bibitem{Shannon1948}
C.~E. Shannon, ``A mathematical theory of communication,'' {\em The Bell System
  Technical Journal}, vol.~27, no.~3, pp.~379--423, 1948.

\bibitem{terhal2002entanglement}
B.~M. Terhal, M.~Horodecki, D.~W. Leung, and D.~P. DiVincenzo, ``The
  entanglement of purification,'' {\em Journal of Mathematical Physics},
  vol.~43, no.~9, pp.~4286--4298, 2002.

\bibitem{chen2012non}
J.~Chen and A.~Winter, ``Non-additivity of the entanglement of purification
  (beyond reasonable doubt),'' {\em arXiv preprint arXiv:1206.1307}, 2012.

\bibitem{dewolfe2020multipartite}
O.~DeWolfe, J.~Levin, and G.~Smith, ``Multipartite optimized correlation
  measures and holography,'' {\em Physical Review D}, vol.~102, no.~6,
  p.~066001, 2020.

\bibitem{lieb1973proof}
E.~H. Lieb and M.~B. Ruskai, ``Proof of the strong subadditivity of
  quantum-mechanical entropy,'' {\em Journal of Mathematical Physics}, vol.~14,
  no.~12, pp.~1938--1941, 1973.

\bibitem{walter2017multipartite}
M.~Walter, D.~Gross, and J.~Eisert, ``Multi-partite entanglement,'' 2017.

\bibitem{komar_quantum_2014}
P.~Kómár, E.~M. Kessler, M.~Bishof, L.~Jiang, A.~S. Sørensen, J.~Ye, and
  M.~D. Lukin, ``A quantum network of clocks,'' {\em Nature Physics}, vol.~10,
  pp.~582--587, Aug. 2014.

\bibitem{PhysRevA.97.042337}
Z.~Eldredge, M.~Foss-Feig, J.~A. Gross, S.~L. Rolston, and A.~V. Gorshkov,
  ``Optimal and secure measurement protocols for quantum sensor networks,''
  {\em Phys. Rev. A}, vol.~97, p.~042337, Apr 2018.

\bibitem{RevModPhys.80.517}
L.~Amico, R.~Fazio, A.~Osterloh, and V.~Vedral, ``Entanglement in many-body
  systems,'' {\em Rev. Mod. Phys.}, vol.~80, pp.~517--576, May 2008.

\bibitem{zeng2018quantum}
B.~Zeng, X.~Chen, D.-L. Zhou, and X.-G. Wen, ``Quantum information meets
  quantum matter -- from quantum entanglement to topological phase in many-body
  systems,'' 2018.

\bibitem{doi:10.1126/science.1247715}
J.~Tura, R.~Augusiak, A.~B. Sainz, T.~Vértesi, M.~Lewenstein, and A.~Acín,
  ``Detecting nonlocality in many-body quantum states,'' {\em Science},
  vol.~344, no.~6189, pp.~1256--1258, 2014.

\bibitem{10.1119/1.16503}
N.~D. Mermin, ``{Quantum mysteries revisited},'' {\em American Journal of
  Physics}, vol.~58, pp.~731--734, 08 1990.

\bibitem{PhysRevLett.106.020405}
J.-D. Bancal, N.~Brunner, N.~Gisin, and Y.-C. Liang, ``Detecting genuine
  multipartite quantum nonlocality: A simple approach and generalization to
  arbitrary dimensions,'' {\em Phys. Rev. Lett.}, vol.~106, p.~020405, Jan
  2011.

\bibitem{Hillery_1999}
M.~Hillery, V.~Bužek, and A.~Berthiaume, ``Quantum secret sharing,'' {\em
  Physical Review A}, vol.~59, p.~1829–1834, Mar. 1999.

\bibitem{PhysRevLett.132.080201}
Y.~Li, Y.~Xiang, X.-D. Yu, H.~C. Nguyen, O.~G\"uhne, and Q.~He, ``Randomness
  certification from multipartite quantum steering for arbitrary dimensional
  systems,'' {\em Phys. Rev. Lett.}, vol.~132, p.~080201, Feb 2024.

\bibitem{PhysRevA.59.4249}
J.~I. Cirac, A.~K. Ekert, S.~F. Huelga, and C.~Macchiavello, ``Distributed
  quantum computation over noisy channels,'' {\em Phys. Rev. A}, vol.~59,
  pp.~4249--4254, Jun 1999.

\bibitem{CALEFFI2024110672}
M.~Caleffi, M.~Amoretti, D.~Ferrari, J.~Illiano, A.~Manzalini, and A.~S.
  Cacciapuoti, ``Distributed quantum computing: A survey,'' {\em Computer
  Networks}, vol.~254, p.~110672, 2024.

\bibitem{D_r_2000}
W.~Dür, G.~Vidal, and J.~I. Cirac, ``Three qubits can be entangled in two
  inequivalent ways,'' {\em Physical Review A}, vol.~62, Nov. 2000.

\bibitem{Verstraete_2002}
F.~Verstraete, J.~Dehaene, B.~De~Moor, and H.~Verschelde, ``Four qubits can be
  entangled in nine different ways,'' {\em Physical Review A}, vol.~65, Apr.
  2002.

\bibitem{christandl2004squashed}
M.~Christandl and A.~Winter, ``Squashed entanglement: An additive entanglement
  measure,'' {\em Journal of mathematical physics}, vol.~45, no.~3,
  pp.~829--840, 2004.

\bibitem{li2021automated}
C.~T. Li, ``An automated theorem proving framework for information-theoretic
  results,'' {\em arXiv preprint arXiv:2101.12370}, 2021.

\end{thebibliography}
\newpage
\appendix
\section{Table of Tripartite Optimized Correlation Measures}
\begin{table}[h]
\begin{center}
\setlength{\extrarowheight}{.15em}
\begin{tabular}{|c|c|c|c|c|}\hline
\begin{tabular}{c}Correlation\cr measure \end{tabular}&\begin{tabular}{c}As symmetrized\cr entropies \end{tabular}& As entropies & As MIs and CMIs & \begin{tabular}{c} Bipartite \cr reduction\end{tabular} \cr \hline\hline
	$\sum I(A_i:A_j) $  &$2 {\cal S}_A - {\cal S}_{AB} $ & \begin{tabular}{c} $2S(A) + 2S(B) + 2S(C)$ \cr $-S(AB) - S(AC) - S(BC)$ \end{tabular} &\begin{tabular}{c}$I(A:B) + I(A:C)$ \cr $ + I(B:C)$ \end{tabular}&  $I(A:B)$\cr\hline
	$J(A:B:C)$  & ${\cal S}_{AB} - 2 {\cal S}_{ABC}$& \begin{tabular}{c}$S(AB) + S(AC) + S(BC)$ \cr $- 2 S(ABC)$ \end{tabular}& $ I(AB:C) + I(A:B|C)$ & $I(A:B)$ \cr\hline \hline
   $f_{c_1} = f^P_3$ &${\cal S}_{Aa}$& $S(Aa) + S(Bb)+ S(Cc)$&\begin{tabular}{c}$I(Aa:Bb) + I(Aa:Cc)$ \cr $ + I(Bb:Cc)$\end{tabular} &  $2 f^P$\cr\hline
\begin{tabular}{c}$f_{c_2}$  \cr (part  of $f^R_3$) \end{tabular}	 & \begin{tabular}{c}${\cal S}_{Aa}  +  2 {\cal S}_{ABC}$ \cr $ - {\cal S}_{ab}  $ \end{tabular}&  \begin{tabular}{c} $S(Aa) + S(Bb) + S(Cc)$\cr $ + 2S(ABC)$ \cr $ - S(ab) - S(ac) - S(bc)$ \end{tabular} &\begin{tabular}{c}$I(a:BC|A) + I(b:AC|B)$ \cr $ + I(c:AB|C)$ \cr $+ I(A:B) + I(AB:C)$ \end{tabular}  & $2 f^R$ \cr\hline
\begin{tabular}{c}$f_{c_3}$  \cr (part  of $f^R_3$) \end{tabular} &\begin{tabular}{c}${\cal S}_{Aa}  + {\cal S}_{ab}$\cr $ - 2 {\cal S}_a  $ \end{tabular}&  \begin{tabular}{c} $S(Aa) + S(Bb) + S(Cc)$ \cr $+ S(ab) + S(ac) + S(bc) $ \cr $-2S(a) - 2 S(b) - 2 S(c)$\end{tabular} &\begin{tabular}{c}$I(a:B|b) + I(a:C|c)$ \cr $ + I(b:A|a)+ I(b:C|c)$ \cr $ + I(c:A|a) + I(c:B|b)$\cr $+ I(A:B|ab) + I(A:C|ac)$ \cr $ + I(B:C|bc) $ \end{tabular} & $2 f^R$ \cr\hline \hline
$f_{s_1} = f^Q_3$ &\begin{tabular}{c}${\cal S}_{Aa}  + {\cal S}_A $ \cr $ + {\cal S}_{AB}- {\cal S}_{ABc}$ \end{tabular}& \begin{tabular}{c} $S(Aa) + S(Bb) + S(Cc)$ \cr $+ S(A)  + S(B) + S(C) $ \cr $+ S(AB) + S(AC) + S(BC)$ \cr $-S(ABc) - S(ACb) - S(BCa)$ \end{tabular}  &\begin{tabular}{c}$I(AB:c) + I(AC:b)$ \cr $ + I(BC:a)+ I(A:BCbc)$ \cr $ + I(B:ACac) + I(C:ABab)$ \end{tabular} &  $4 f^Q$ \cr\hline
  $f_{s_2}$	 & \begin{tabular}{c}$2{\cal S}_{Aa}  + {\cal S}_{AB}$ \cr $ - {\cal S}_{ABa}$ \end{tabular}& \begin{tabular}{c} $2S(Aa) + 2S(Bb)+ 2S(Cc)$ \cr $+S(AB) + S(AC) + S(BC)$ \cr $-S(ABa) - S(ABb) - S(ACa)$ \cr $ - S(ACc) - S(BCb) - S(BCc)$ \end{tabular} &\begin{tabular}{c}$I(A:b|B) + I(A:Cc) $ \cr $+ I(B:c|C) + I(B:Aa)$ \cr $ + I(C:a|A) + I(C:Bb)$ \end{tabular} &  $ 2 f^R$ \cr\hline\hline

	$f_{a_1}$  & \begin{tabular}{c}$2 {\cal S}_A + {\cal S}_{ABa}$ \cr $ - {\cal S}_{Ab}$\end{tabular} &\begin{tabular}{c} $2S(A) + 2S(B) + 2S(C)$ \cr $+S(ABa) + S(ABb) + S(ACa)$\cr $ + S(ACc) + S(BCb) + S(BCc)$\cr $-S(Ab) - S(Ba) - S(Ac) $\cr $- S(Ca) - S(Bc) - S(Cb)$ \end{tabular} &\begin{tabular}{c}$I(A : Cbc) + I(A: Bbc)$ \cr $ + I(B:Cac)+ I(B:Aac)$\cr $ + I(C:Bab) + I(C : Aab)$ \end{tabular} &   $2f^Q$\cr\hline
$f_{a_2}$	 &\begin{tabular}{c} $2{\cal S}_{Aa}  + 2 {\cal S}_A $\cr $+ {\cal S}_{ABc}$ \cr $- {\cal S}_{Ab}- {\cal S}_a   $\end{tabular} & \begin{tabular}{c} $2S(Aa) + 2S(Bb) + 2S(Cc)$ \cr $+ 2 S(A) + 2 S(B) + 2 S(C)$\cr $+S(ABc) + S(ACb) + S(BCa)$ \cr $-S(Ab) - S(Ba) - S(Ac) $\cr $- S(Ca) - S(Bc) - S(Cb)$\cr $-S(a) - S(b)- S(c)$\end{tabular}  &\begin{tabular}{c}$I(A:Bc|a) + I(A:Cb|a)$ \cr $ + I(B:Ac|b)+ I(B:Ca|b)$ \cr $ + I(C:Ab|c) + I(C:Ba|c)$ \cr $+ I(A:B) + I(A:C)$ \cr $ + I(B:C)+ I(AB:c) $ \cr $+ I(AC:b) + I(BC:a)$ \end{tabular} &  \begin{tabular}{c}$4 f^R+$ \cr $  I(A:B)$ \end{tabular}\cr\hline\hline
$f_{r_1}$	 &\begin{tabular}{c}${\cal S}_{Aa} $ \cr $ + {\cal S}_A + {\cal S}_{Ab}$ \cr $ - {\cal S}_{ABc} - 2 {\cal S}_a  $ \end{tabular}& \begin{tabular}{c} $S(Aa) + S(Bb) + S(Cc)$ \cr $+S(A) + S(B) + S(C)$ \cr $+S(Ab) + S(Ba) + S(Ac)$\cr $ + S(Ca) + S(Bc) + S(Cb)$ \cr $-S(ABc) - S(ACb) - S(BCa)$ \cr $-2S(a) - 2 S(b) - 2 S(c)$\end{tabular}  &\begin{tabular}{c}$I(A:B|c) + I(A:C|b)$ \cr $ + I(B:C|a)+ I(A:BCbc)$ \cr $ + I(B:ACac) + I(C:ABab)$ \end{tabular} & \begin{tabular}{c}$2 f^R +$ \cr $ 2I(A:B)$ \end{tabular} \cr\hline \hline
$f_{b_1} = \tilde{f}^{\rm sq}_3$	 & \begin{tabular}{c}${\cal S}_{Aa}  + {\cal S}_{ABc}$\cr $ - {\cal S}_{ab} - {\cal S}_a  $ \end{tabular}& \begin{tabular}{c} $S(Aa) + S(Bb) + S(Cc)$ \cr $+S(ABc) + S(ACb) + S(BCa)$\cr $- S(ab) - S(ac) - S(bc)$ \cr $- S(a) - S(b) - S(c)$ \end{tabular}  &\begin{tabular}{c}$I(A:BC|a) + I(B:AC|b)$\cr $ + I(C:AB|c)$\end{tabular}  &$4 f^{\rm sq}$\cr\hline
 $f_{b_2}$   & \begin{tabular}{c}${\cal S}_{ABa}$ \cr $ - {\cal S}_{ab}-{\cal S}_{Aa}$\end{tabular} & \begin{tabular}{c}$S(ABa) + S(ABb) + S(ACa)$ \cr $ + S(ACc) + S(BCb) + S(BCc)$\cr $ - S(ab) - S(ac) - S(bc)$ \cr $ - S(Aa) - S(Bb)- S(Cc)$ \end{tabular} &\begin{tabular}{c}$I(B:C|Aa) + I(C:A|Bb)$ \cr $ + I(A:B|Cc)$\end{tabular}&  $ I(A:B)$ \cr\hline
 $f_{b_3}$ &\begin{tabular}{c}$2{\cal S}_{Aa}  + {\cal S}_{Ab}$\cr $ - {\cal S}_{ABa} - 2{\cal S}_a $\end{tabular} & \begin{tabular}{c} $2S(Aa) + 2S(Bb)+ 2S(Cc)$ \cr $+S(Ab) + S(Ba) + S(Ac)$ \cr $+S(Ca) + S(Bc) + S(Cb)$ \cr $-S(ABa) - S(ABb) - S(ACa)$ \cr $ - S(ACc) - S(BCb) - S(BCc)$ \cr $-2 S(a)- 2 S(b) - 2 S(c)$\end{tabular}&\begin{tabular}{c}$I(A:B|b) + I(A:C|c)$ \cr  $+ I(B:A|a)+ I(B:C|c)$ \cr $ + I(C:A|a)+ I(C:B|b)$ \end{tabular}  & $4 f^{\rm sq}$ \cr\hline
	\end{tabular}	
\end{center}
\caption{Tripartite optimized correlation measures from \cite{dewolfe2020multipartite}.   \label{tripartite OCMs}}
\end{table}
\section*{Entropy Inequalities}\label{Entropy Ineqs}
\subsection{Strong Subadditivity}
\emph{Strong subadditivity} of the von Neumann entropy, proved in \cite{lieb1973proof}, is an inequality which states that quantum conditional mutual information cannot be negative.  More precisely, given any tripartite state $\rho_{ABC}$,
\begin{align}
    I(A:B|C)_{\rho}\equiv S(\rho_{AC}) + S(\rho_{BC}) - S(\rho_{ABC}) - S(\rho_{C}) \geq 0
\end{align}
must hold.  This inequality, along with several equivalent versions shown below, is strong enough to establish all additivities found in this paper as well as in \cite{cross2017uniform}.
\subsection{Inequalities Equivalent to Strong Subadditivity}
\emph{Weak monotonicity} of the von Neumann entropy is the inequality
\begin{align}
    S(A|B) + S(A|C)\geq 0,
\end{align}
which holds for all tripartite states $\rho_{ABC}$. This can be shown to be equivalent to strong subadditivity by first purifying $\rho_{ABC}$ to $\rho_{ABCD}$, and then rewriting the conditional mutual information as
\begin{align}
    I(A:B|C)_{\rho} = S(\rho_{AC}) + S(\rho_{AD}) - S(\rho_{D}) - S(\rho_{C}) = S(A|C)_\rho + S(A|D)_\rho.
\end{align}

\section*{Equivalence classes of standard decouplings for tripartite formulas with two ancillas}

Here we consider the case with 3 parties $A$, $B$, and $C$, and 2 ancillas $V$ and $W$ (and a purifying system $X$).  Decouplings in this case consist of two ordered pairs $(a,b)(c,d)$, where $a$ and $c$ are disjoint, and $b$ and $d$ are disjoint (this is the consistency requirement). The ordered pair $(a,b)$ gives a recipe for $V_1$ and $V_2$, while $(c,d)$ gives a recipe for $W_1$ and $W_2$, (we have slightly modified the notation of \cite{cross2017uniform} for clarity).  The decouplings have the following three symmetries:

\begin{enumerate}
    \item The 1 and 2 systems can be swapped: $(a,b)(c,d) \to (b,a)(d,c)$
    \item $A$, $B$, and $C$ can be permuted, inducing a permutation $P$ of $\mathcal{P}(\{A,B,C\})$: $(a,b)(c,d) \to (P(a),P(b))(P(c),P(d))$
    \item $V$, $W$, and $X$ can be permuted
\end{enumerate}

These symmetries generate the following 22 equivalence classes of consistent standard decouplings for 3 parties and 2 ancillas:
\begin{enumerate}
\begin{multicols}{4}
    
    \item $\begin{Bmatrix}
    \bm{(0,0)(0,0)}\\
    (0,0)(7,7)\\
    (7,7)(0,0)
    \end{Bmatrix}$
    
    \item $\begin{Bmatrix}
    \bm{(0,0)(0,7)}\\
    (0,7)(0,0)\\
    (7,0)(0,7)\\
    (0,0)(7,0)\\
    (0,7)(7,0)\\
    (7,0)(0,0)
    \end{Bmatrix}$

    \item $\begin{Bmatrix}
    \bm{(1,1)(2,2)}\\
    \bm{(2,2)(1,1)}\\
    \bm{(3,3)(2,2)}\\
    \bm{(1,1)(3,3)}\\
    \bm{(2,2)(3,3)}\\
    \bm{(3,3)(1,1)}
    \end{Bmatrix}$
    
    \item $\begin{Bmatrix}
    \bm{(1,2)(2,3)} & (2,3)(1,2)\\
    \bm{(2,1)(1,3)} & (1,3)(2,1)\\
    \bm{(3,2)(2,1)} & (2,1)(3,2)\\
    \bm{(1,3)(3,2)} & (3,2)(1,3)\\
    \bm{(2,3)(3,1)} & (3,1)(2,3)\\
    \bm{(3,1)(1,2)} & (1,2)(3,1)
    \end{Bmatrix}$

\end{multicols}
\begin{multicols}{2}

    \item $\begin{Bmatrix}
    \bm{(1,1)(0,0)} & (0,0)(1,1) & (1,1)(6,6)\\
    \bm{(2,2)(0,0)} & (0,0)(2,2) & (2,2)(5,5)\\
    \bm{(3,3)(0,0)} & (0,0)(3,3) & (3,3)(4,4)\\
    (4,4)(0,0) & (0,0)(4,4) & (4,4)(3,3)\\
    (5,5)(0,0) & (0,0)(5,5) & (5,5)(2,2)\\
    (6,6)(0,0) & (0,0)(6,6) & (6,6)(1,1)
    \end{Bmatrix}$
    \newline
    \newline
    \item $\begin{Bmatrix}
    \bm{(4,4)(3,0)} & (3,0)(4,4) & (0,3)(3,0)\\
    \bm{(6,6)(1,0)} & (1,0)(6,6) & (0,1)(1,0)\\
    \bm{(5,5)(2,0)} & (2,0)(5,5) & (0,2)(2,0)\\
    (4,4)(0,3) & (0,3)(4,4) & (3,0)(0,3)\\
    (6,6)(0,1) & (0,1)(6,6) & (1,0)(0,1)\\
    (5,5)(0,2) & (0,2)(5,5) & (2,0)(0,2)
    \end{Bmatrix}$
    \newline
    \newline
    \item $\begin{Bmatrix}
    \bm{(1,1)(6,0)} & (6,0)(1,1) & (0,6)(6,0)\\
    \bm{(2,2)(5,0)} & (5,0)(2,2) & (0,5)(5,0)\\
    \bm{(3,3)(4,0)} & (4,0)(3,3) & (0,4)(4,0)\\
    (1,1)(0,6) & (0,6)(1,1) & (6,0)(0,6)\\
    (2,2)(0,5) & (0,5)(2,2) & (5,0)(0,5)\\
    (3,3)(0,4) & (0,4)(3,3) & (4,0)(0,4)
    \end{Bmatrix}$
    \newline
    \newline
    \item $\begin{Bmatrix}
    \bm{(1,1)(2,3)} & (2,3)(1,1) & (3,2)(2,3)\\
    \bm{(2,2)(1,3)} & (1,3)(2,2) & (3,1)(1,3)\\
    \bm{(3,3)(2,1)} & (2,1)(3,3) & (1,2)(2,1)\\
    \bm{(1,1)(3,2)} & (3,2)(1,1) & (2,3)(3,2)\\
    \bm{(2,2)(3,1)} & (3,1)(2,2) & (1,3)(3,1)\\
    \bm{(3,3)(1,2)} & (1,2)(3,3) & (2,1)(1,2)
    \end{Bmatrix}$
    \newline
    \newline
    \item $\begin{Bmatrix}
    \bm{(0,0)(1,6)} & (1,6)(0,0) & (6,1)(1,6)\\
    \bm{(0,0)(2,5)} & (2,5)(0,0) & (5,2)(2,5)\\
    \bm{(0,0)(3,4)} & (3,4)(0,0) & (4,3)(3,4)\\
    (0,0)(6,1) & (6,1)(0,0) & (1,6)(6,1)\\
    (0,0)(5,2) & (5,2)(0,0) & (2,5)(5,2)\\
    (0,0)(4,3) & (4,3)(0,0) & (3,4)(4,3)
    \end{Bmatrix}$
    
\end{multicols}

\item $\begin{Bmatrix}
\bm{(1,0)(0,0)} & (0,0)(1,0) & (6,7)(0,0) & (1,0)(6,7) & (6,7)(1,0) & (0,0)(6,7)\\
\bm{(2,0)(0,0)} & (0,0)(2,0) & (5,7)(0,0) & (2,0)(5,7) & (5,7)(2,0) & (0,0)(5,7)\\
\bm{(3,0)(0,0)} & (0,0)(3,0) & (4,7)(0,0) & (3,0)(4,7) & (4,7)(3,0) & (0,0)(4,7)
\end{Bmatrix}\cup$ $\{$symmetry 1$\}$
\newline
\newline
\item $\begin{Bmatrix}
\bm{(1,0)(2,0)} & \bm{(2,0)(1,0)} & \bm{(1,0)(3,0)} & \bm{(3,0)(1,0)} & \bm{(3,0)(2,0)} & \bm{(2,0)(3,0)}\\
(3,7)(2,0) & (3,7)(1,0) & (2,7)(3,0) & (2,7)(1,0) & (1,7)(2,0) & (1,7)(3,0)\\
(1,0)(3,7) & (2,0)(3,7) & (1,0)(2,7) & (3,0)(2,7) & (3,0)(1,7) & (2,0)(1,7)
\end{Bmatrix}\cup$ $\{$symmetry 1$\}$
\newline
\newline
\item $\begin{Bmatrix}
\bm{(1,0)(6,0)} & (6,0)(1,0) & (0,7)(6,0) & (1,0)(0,7) & (0,7)(1,0) & (6,0)(0,7)\\
\bm{(2,0)(5,0)} & (5,0)(2,0) & (0,7)(5,0) & (2,0)(0,7) & (0,7)(2,0) & (5,0)(0,7)\\
\bm{(3,0)(4,0)} & (4,0)(3,0) & (0,7)(4,0) & (3,0)(0,7) & (0,7)(3,0) & (4,0)(0,7)
\end{Bmatrix}\cup$ $\{$symmetry 1$\}$
\newline
\newline
\item $\begin{Bmatrix}
\bm{(1,7)(6,0)} & (6,0)(1,7) & (0,0)(6,0) & (1,7)(0,0) & (0,0)(1,7) & (6,0)(0,0)\\
\bm{(2,7)(5,0)} & (5,0)(2,7) & (0,0)(5,0) & (2,7)(0,0) & (0,0)(2,7) & (5,0)(0,0)\\
\bm{(3,7)(4,0)} & (4,0)(3,7) & (0,0)(4,0) & (3,7)(0,0) & (0,0)(3,7) & (4,0)(0,0)
\end{Bmatrix}\cup$ $\{$symmetry 1$\}$
\newline
\newline
\item $\begin{Bmatrix}
\bm{(0,1)(1,6)} & (1,6)(0,1) & (6,0)(1,6) & (0,1)(6,0) & (6,0)(0,1) & (1,6)(6,0)\\
\bm{(0,2)(2,5)} & (2,5)(0,2) & (5,0)(2,5) & (0,2)(5,0) & (5,0)(0,2) & (2,5)(5,0)\\
\bm{(0,3)(3,4)} & (3,4)(0,3) & (4,0)(3,4) & (0,3)(4,0) & (4,0)(0,3) & (3,4)(4,0)
\end{Bmatrix}\cup$ $\{$symmetry 1$\}$
\newline
\newline
\item $\begin{Bmatrix}
\bm{(1,2)(0,0)} & (0,0)(1,2) & (6,5)(0,0) & (1,2)(6,5) & (6,5)(1,2) & (0,0)(6,5)\\
\bm{(2,1)(0,0)} & (0,0)(2,1) & (5,6)(0,0) & (2,1)(5,6) & (5,6)(2,1) & (0,0)(5,6)\\
\bm{(3,2)(0,0)} & (0,0)(3,2) & (4,5)(0,0) & (3,2)(4,5) & (4,5)(3,2) & (0,0)(4,5)\\
\bm{(2,3)(0,0)} & (0,0)(2,3) & (5,4)(0,0) & (2,3)(5,4) & (5,4)(2,3) & (0,0)(5,4)\\
\bm{(1,3)(0,0)} & (0,0)(1,3) & (6,4)(0,0) & (1,3)(6,4) & (6,4)(1,3) & (0,0)(6,4)\\
\bm{(3,1)(0,0)} & (0,0)(3,1) & (4,6)(0,0) & (3,1)(4,6) & (4,6)(3,1) & (0,0)(4,6)
\end{Bmatrix}$
\newline
\newline
\item $\begin{Bmatrix}
\bm{(1,4)(0,0)} & (0,0)(1,4) & (6,3)(0,0) & (1,4)(6,3) & (6,3)(1,4) & (0,0)(6,3)\\
(4,1)(0,0) & (0,0)(4,1) & \bm{(3,6)(0,0)} & (4,1)(3,6) & (3,6)(4,1) & (0,0)(3,6)\\
\bm{(2,4)(0,0)} & (0,0)(2,4) & (5,3)(0,0) & (2,4)(5,3) & (5,3)(2,4) & (0,0)(5,3)\\
(4,2)(0,0) & (0,0)(4,2) & \bm{(3,5)(0,0)} & (4,2)(3,5) & (3,5)(4,2) & (0,0)(3,5)\\
\bm{(1,5)(0,0)} & (0,0)(1,5) & (6,2)(0,0) & (1,5)(6,2) & (6,2)(1,5) & (0,0)(6,2)\\
(5,1)(0,0) & (0,0)(5,1) & \bm{(2,6)(0,0)} & (5,1)(2,6) & (2,6)(5,1) & (0,0)(2,6)
\end{Bmatrix}$
\newline
\newline
\item $\begin{Bmatrix}
\bm{(1,2)(6,0)} & (6,0)(1,2) & (0,5)(6,0) & (1,2)(0,5) & (0,5)(1,2) & (6,0)(0,5)\\
(2,1)(0,6) & (0,6)(2,1) & (5,0)(0,6) & \bm{(2,1)(5,0)} & (5,0)(2,1) & (0,6)(5,0)\\
\bm{(3,2)(4,0)} & (4,0)(3,2) & (0,5)(4,0) & (3,2)(0,5) & (0,5)(3,2) & (4,0)(0,5)\\
(2,3)(0,4) & (0,4)(2,3) & (5,0)(0,4) & \bm{(2,3)(5,0)} & (5,0)(2,3) & (0,4)(5,0)\\
\bm{(1,3)(6,0)} & (6,0)(1,3) & (0,4)(6,0) & (1,3)(0,4) & (0,4)(1,3) & (6,0)(0,4)\\
(3,1)(0,6) & (0,6)(3,1) & (4,0)(0,6) & \bm{(3,1)(4,0)} & (4,0)(3,1) & (0,6)(4,0)
\end{Bmatrix}$
\newline
\newline
\item $\begin{Bmatrix}
\bm{(4,5)(3,0)} & (3,0)(4,5) & (0,2)(3,0) & (4,5)(0,2) & (0,2)(4,5) & (3,0)(0,2)\\
(5,4)(0,3) & (0,3)(5,4) & (2,0)(0,3) & \bm{(5,4)(2,0)} & (2,0)(5,4) & (0,3)(2,0)\\
\bm{(4,6)(3,0)} & (3,0)(4,6) & (0,1)(3,0) & (4,6)(0,1) & (0,1)(4,6) & (3,0)(0,1)\\
(6,4)(0,3) & (0,3)(6,4) & (1,0)(0,3) & \bm{(6,4)(1,0)} & (1,0)(6,4) & (0,3)(1,0)\\
\bm{(6,5)(1,0)} & (1,0)(6,5) & (0,2)(1,0) & (6,5)(0,2) & (0,2)(6,5) & (1,0)(0,2)\\
(5,6)(0,1) & (0,1)(5,6) & (2,0)(0,1) & \bm{(5,6)(2,0)} & (2,0)(5,6) & (0,1)(2,0)
\end{Bmatrix}$
\newline
\newline
\item $\begin{Bmatrix}
\bm{(1,2)(2,0)} & (2,0)(1,2) & (3,5)(2,0) & (1,2)(3,5) & (3,5)(1,2) & (2,0)(3,5)\\
\bm{(2,1)(1,0)} & (1,0)(2,1) & (3,6)(1,0) & (2,1)(3,6) & (3,6)(2,1) & (1,0)(3,6)\\
\bm{(3,2)(2,0)} & (2,0)(3,2) & (1,5)(2,0) & (3,2)(1,5) & (1,5)(3,2) & (2,0)(1,5)\\
\bm{(1,3)(3,0)} & (3,0)(1,3) & (2,4)(3,0) & (1,3)(2,4) & (2,4)(1,3) & (3,0)(2,4)\\
\bm{(2,3)(3,0)} & (3,0)(2,3) & (1,4)(3,0) & (2,3)(1,4) & (1,4)(2,3) & (3,0)(1,4)\\
\bm{(3,1)(1,0)} & (1,0)(3,1) & (2,6)(1,0) & (3,1)(2,6) & (2,6)(3,1) & (1,0)(2,6)
\end{Bmatrix}\cup$ $\{$symmetry 1$\}$
\newline
\newline
\item $\begin{Bmatrix}
\bm{(1,4)(2,0)} & (2,0)(1,4) & (3,3)(2,0) & (1,4)(3,3) & (3,3)(1,4) & (2,0)(3,3)\\
\bm{(2,4)(1,0)} & (1,0)(2,4) & (3,3)(1,0) & (2,4)(3,3) & (3,3)(2,4) & (1,0)(3,3)\\
\bm{(3,6)(2,0)} & (2,0)(3,6) & (1,1)(2,0) & (3,6)(1,1) & (1,1)(3,6) & (2,0)(1,1)\\
\bm{(1,5)(3,0)} & (3,0)(1,5) & (2,2)(3,0) & (1,5)(2,2) & (2,2)(1,5) & (3,0)(2,2)\\
\bm{(2,6)(3,0)} & (3,0)(2,6) & (1,1)(3,0) & (2,6)(1,1) & (1,1)(2,6) & (3,0)(1,1)\\
\bm{(3,5)(1,0)} & (1,0)(3,5) & (2,2)(1,0) & (3,5)(2,2) & (2,2)(3,5) & (1,0)(2,2)
\end{Bmatrix}\cup$ $\{$symmetry 1$\}$
\newline
\newline
\item $\begin{Bmatrix}
\bm{(1,4)(6,0)} & (6,0)(1,4) & (0,3)(6,0) & (1,4)(0,3) & (0,3)(1,4) & (6,0)(0,3)\\
\bm{(2,4)(5,0)} & (5,0)(2,4) & (0,3)(5,0) & (2,4)(0,3) & (0,3)(2,4) & (5,0)(0,3)\\
\bm{(3,6)(4,0)} & (4,0)(3,6) & (0,1)(4,0) & (3,6)(0,1) & (0,1)(3,6) & (4,0)(0,1)\\
\bm{(1,5)(6,0)} & (6,0)(1,5) & (0,2)(6,0) & (1,5)(0,2) & (0,2)(1,5) & (6,0)(0,2)\\
\bm{(2,6)(5,0)} & (5,0)(2,6) & (0,1)(5,0) & (2,6)(0,1) & (0,1)(2,6) & (5,0)(0,1)\\
\bm{(3,5)(4,0)} & (4,0)(3,5) & (0,2)(4,0) & (3,5)(0,2) & (0,2)(3,5) & (4,0)(0,2)
\end{Bmatrix}\cup$ $\{$symmetry 1$\}$
\newline
\newline
\item $\begin{Bmatrix}
\bm{(1,2)(6,1)} & (6,1)(1,2) & (0,3)(6,1) & (1,2)(0,3) & (0,3)(1,2) & (6,1)(0,3)\\
\bm{(2,1)(5,2)} & (5,2)(2,1) & (0,3)(5,2) & (2,1)(0,3) & (0,3)(2,1) & (5,2)(0,3)\\
\bm{(3,2)(4,3)} & (4,3)(3,2) & (0,1)(4,3) & (3,2)(0,1) & (0,1)(3,2) & (4,3)(0,1)\\
\bm{(1,3)(6,1)} & (6,1)(1,3) & (0,2)(6,1) & (1,3)(0,2) & (0,2)(1,3) & (6,1)(0,2)\\
\bm{(2,3)(5,2)} & (5,2)(2,3) & (0,1)(5,2) & (2,3)(0,1) & (0,1)(2,3) & (5,2)(0,1)\\
\bm{(3,1)(4,3)} & (4,3)(3,1) & (0,2)(4,3) & (3,1)(0,2) & (0,2)(3,1) & (4,3)(0,2)
\end{Bmatrix}\cup$ $\{$symmetry 1$\}$
\end{enumerate}
\vspace{10pt}
According to Theorem G.1 of \cite{cross2017uniform}, the $(a,b)(c,d)$ cone is a direct sum of the $(a,b)$, $(c,d)$, and $(a\cup c,b\cup d)$ cones.  The $(a,b)$ term of this direct sum corresponds to subsets of $\{A,B,C,V,W\}$ that contain $V$ and not $W$, the $(c,d)$ term $W$ and not $V$, and the $(a\cup c,b\cup d)$ term both $V$ and $W$.

\section*{Uniform Additivity Cones of Tripartite Formulas with 1 Ancilla}

In Sec. \ref{Tripartite 1ancilla cones}, we derived the uniform additivity cone for the $(0,0)$ decoupling. Here we derive the uniform additivity cones for the remaining seven equivalence classes.
\begin{enumerate}
   \item $(7,0)$ decoupling:
    \begin{multline}
        \Delta^{V,(7,0)} = -\alpha_{AV}I(A_1:B_2C_2|A_2V) - \alpha_{BV}I(B_1:A_2C_2|B_2V) - \alpha_{CV}I(C_1:A_2B_2|C_2V)\\ - \alpha_{ABV}I(A_1B_1:C_2|A_2B_2V) - \alpha_{ACV}I(A_1C_1:B_2|A_2C_2V) - \alpha_{BCV}I(B_1C_1:A_2|B_2C_2V)
        \nonumber
    \end{multline}
    \subitem \textit{Necessary conditions:} The conditions
    \begin{subequations}
    \begin{align}
        \alpha_{AV} &\geq 0\label{2-17}\\
        \alpha_{BV} &\geq 0\label{2-18}\\
        \alpha_{CV} &\geq 0\label{2-19}\\
        \alpha_{ABV} &\geq 0\label{2-14}\\
        \alpha_{ACV} &\geq 0\label{2-15}\\
        \alpha_{BCV} &\geq 0\label{2-16}
    \end{align}\label{ineq_70}
    \end{subequations}
    are necessary for non-positivity of $\Delta^{V,(7,0)}$. These inequalities are implied by the following states:
    \begin{subequations}
    \begin{align}
    &A_1 = \mathcal{B}_1&&A_2 = \mathcal{B}_2, B_2 = C_2 = \mathcal{B}_3&&&V = \mathcal{B}_1\oplus\mathcal{B}_2\oplus\mathcal{B}_3
    \\&B_1 = \mathcal{B}_1&&B_2 = \mathcal{B}_2, A_2 = C_2 = \mathcal{B}_3&&&V = \mathcal{B}_1\oplus\mathcal{B}_2\oplus\mathcal{B}_3
    \\&C_1 = \mathcal{B}_1&&C_2 = \mathcal{B}_2, A_2 = B_2 = \mathcal{B}_3&&&V = \mathcal{B}_1\oplus\mathcal{B}_2\oplus\mathcal{B}_3
    \\&A_1 = \mathcal{B}_1, B_1 = \mathcal{B}_2&&C_2 = \mathcal{B}_3&&&V = \mathcal{B}_1\oplus\mathcal{B}_2\oplus\mathcal{B}_3
    \\&A_1 = \mathcal{B}_1, C_1 = \mathcal{B}_2&&B_2 = \mathcal{B}_3&&&V = \mathcal{B}_1\oplus\mathcal{B}_2\oplus\mathcal{B}_3
    \\&B_1 = \mathcal{B}_1, C_1 = \mathcal{B}_2&&A_2 = \mathcal{B}_3&&&V = \mathcal{B}_1\oplus\mathcal{B}_2\oplus\mathcal{B}_3
    \end{align}
    \end{subequations}
    \subitem \textit{Sufficient conditions:}  These are also sufficient.
    
    Applying the balancing condition and dualizing (\ref{ineq_70}), we find a minimal generating set of the (7,0) cone:
     \begin{align}
        (-1,0,0,0,1,0,0,0) &\to S(AB|V) \nonumber\\
        (-1,0,0,0,0,1,0,0) &\to S(AC|V) \nonumber\\
        (-1,0,0,0,0,0,1,0) &\to S(BC|V) \nonumber\\
        (-1,1,0,0,0,0,0,0) &\to S(A|V) \nonumber\\
        (-1,0,1,0,0,0,0,0) &\to S(B|V) \nonumber\\
        (-1,0,0,1,0,0,0,0) &\to S(C|V) \nonumber\\
        (1,0,0,0,0,0,0,-1) &\to -S(ABC|V) \nonumber\\
        (-1,0,0,0,0,0,0,1) &\to S(ABC|V) \nonumber
    \end{align}
    \item $(1,1)$ decoupling:
    \begin{multline}
        \Delta^{V,(1,1)} = \alpha_VI(A_1:A_2|V) + \alpha_{ABV}I(B_1:B_2|A_1A_2V) + \alpha_{ACV}I(C_1:C_2|A_1A_2V)\\ + \alpha_{ABCV}I(B_1C_1:B_2C_2|A_1A_2V)
        + \alpha_{BV}[S(B_1A_2V) + S(A_1B_2V) - S(B_1B_2V) - S(A_1A_2V)]\\
         + \alpha_{CV}[S(C_1A_2V) + S(A_1C_2V) - S(C_1C_2V) - S(A_1A_2V)]\\
         + \alpha_{BCV}[S(B_1C_1A_2V) + S(A_1B_2C_2V) - S(B_1C_1B_2C_2V) - S(A_1A_2V)]\\
         \nonumber
    \end{multline}
    \subitem \textit{Necessary conditions:} The conditions
    \begin{subequations}
    \begin{align}
    -\alpha_{ABCV} - \alpha_{ABV} - \alpha_{ACV} - \alpha_{BCV} - \alpha_{BV} - \alpha_{CV} &\geq 0\\
    -\alpha_{ABCV} - \alpha_{ABV} - \alpha_{BCV} - \alpha_{BV} &\geq 0\\
    -\alpha_{ABCV} - \alpha_{ACV} - \alpha_{BCV} - \alpha_{CV} &\geq 0\\
    -\alpha_{BCV} - \alpha_{BV} - \alpha_{CV} - \alpha_{V} &\geq 0\\
    -\alpha_{ABCV} - \alpha_{BCV} &\geq 0\\
    \alpha_{BCV} &\geq 0\\
    \alpha_{BV} &\geq 0\\
    \alpha_{CV} &\geq 0
    \end{align}
    \end{subequations}
    are necessary for non-positivity of $\Delta^{V,(1,1)}$. These inequalities are implied by the following states:
    \begin{subequations}
    \begin{align}
    &B_1 = C_1 = \mathcal{B}_1&&B_2 = C_2 = \mathcal{B}_2&&&V = \mathcal{B}_1\oplus\mathcal{B}_2
    \\&B_1 = \mathcal{B}_1&&B_2 = \mathcal{B}_2&&&V = \mathcal{B}_1\oplus\mathcal{B}_2
    \\&C_1 = \mathcal{B}_1&&C_2 = \mathcal{B}_2&&&V = \mathcal{B}_1\oplus\mathcal{B}_2
    \\&A_1 = \mathcal{B}_1&&A_2 = \mathcal{B}_2&&&V = \mathcal{B}_1\oplus\mathcal{B}_2
    \\&C_1 = \mathcal{B}_1&&B_2 = \mathcal{B}_2&&&V = \mathcal{B}_1\oplus \mathcal{B}_2
    \\&B_1 = \mathcal{B}_1, C_1 = \mathcal{B}_2&&A_2 = \mathcal{B}_3&&&V = \mathcal{B}_1\oplus\mathcal{B}_2\oplus\mathcal{B}_3
    \\&B_1 = \mathcal{B}_1&&A_2 = C_2 = \mathcal{B}_2&&&V = \mathcal{B}_1\oplus\mathcal{B}_2
    \\&C_1 = \mathcal{B}_1&&A_2 = B_2 = \mathcal{B}_2&&&V = \mathcal{B}_1\oplus\mathcal{B}_2
    \end{align}
    \end{subequations}
    \subitem \textit{Sufficient conditions:}  These conditions are not sufficient for $\Delta^{V,(1,1)}\leq 0$, since there are vectors in the dual of the cone that they span for which this inequality is false. That cone, whose minimal generating set is shown below, is an outer bound for the $(1,1)$ cone. Removing the generators for which $\Delta^{V,(1,1)}\leq 0$ is false, (the three conditional mutual informations at the top of the list) gives an inner bound for the $(1,1)$ cone.
    \begin{align}
        (-1,1,0,0,0,0,1,-1) &\to I(A:BC|V) \nonumber\\
        (-1,1,1,0,-1,0,0,0) &\to I(A:B|V) \nonumber\\
        (-1,1,0,1,0,-1,0,0) &\to I(A:C|V) \nonumber\\
        (-1,1,0,0,0,0,0,0) &\to S(A|V) \nonumber\\
        (0,1,0,0,-1,0,0,0) &\to -S(B|AV) \nonumber\\
        (0,1,0,0,0,-1,0,0) &\to -S(C|AV) \nonumber\\
        (0,0,0,0,0,1,0,-1) &\to -S(B|ACV) \nonumber\\
        (0,0,0,0,1,0,0,-1) &\to -S(C|ABV) \nonumber
    \end{align}
    
    \item $(1,6)$ decoupling:
    \begin{multline}
        \Delta^{V,(1,6)} = \alpha_VI(B_1C_1:A_2|V) + \alpha_{ABCV}I(A_1:B_2C_2|B_1C_1A_2V)\\
        + \alpha_{BV}[S(B_1A_2V) + S(B_1C_1B_2V) - S(B_1B_2V) - S(B_1C_1A_2V)]\\
        + \alpha_{CV}[S(C_1A_2V) + S(B_1C_1C_2V) - S(C_1C_2V) - S(B_1C_1A_2V)]\\
        + \alpha_{ABV}[S(A_1B_1A_2V) + S(B_1C_1A_2B_2V) - S(A_1B_1A_2B_2V) - S(B_1C_1A_2V)]\\
        + \alpha_{ACV}[S(A_1C_1A_2V) + S(B_1C_1A_2C_2V) - S(A_1C_1A_2C_2V) - S(B_1C_1A_2V)]\\
        \nonumber
    \end{multline}
    \subitem \textit{Necessary conditions:} The conditions
    \begin{subequations}
    \begin{align}
    -\alpha_{ABV} - \alpha_{ACV} - \alpha_{ABCV}&\geq 0 \label{4-1}\\
    -\alpha_V - \alpha_{BV} - \alpha_{CV}&\geq 0 \label{4-2}\\
    \alpha_{ABV}&\geq 0 \label{4-3}\\
    \alpha_{ACV}&\geq 0 \label{4-4}\\
    \alpha_{BV}&\geq 0 \label{4-5}\\
    \alpha_{CV}&\geq 0 \label{4-6}
    \end{align}
    \end{subequations}
    are necessary for non-positivity of $\Delta^{V,(1,6)}$. These inequalities are implied by the following states:
    \begin{subequations}
    \begin{align}
    &A_1 = \mathcal{B}_1&&B_2 = C_2 = \mathcal{B}_2&&&V = \mathcal{B}_1\oplus\mathcal{B}_2
    \\&B_1 = \mathcal{B}_1, C_1 = \mathcal{B}_2&&A_2 = \mathcal{B}_3&&&V = \mathcal{B}_1\oplus\mathcal{B}_2\oplus\mathcal{B}_3
    \\&C_1 = \mathcal{B}_1&&A_2 = \mathcal{B}_2, B_2 = \mathcal{B}_3&&&V = \mathcal{B}_1\oplus\mathcal{B}_2\oplus\mathcal{B}_3
    \\&B_1 = \mathcal{B}_1&&A_2 = \mathcal{B}_2, C_2 = \mathcal{B}_3&&&V = \mathcal{B}_1\oplus\mathcal{B}_2\oplus\mathcal{B}_3
    \\&A_1 = C_1 = \mathcal{B}_1&&B_2 = \mathcal{B}_2&&&V = \mathcal{B}_1\oplus \mathcal{B}_2
    \\&A_1 = B_1 = \mathcal{B}_1&&C_2 = \mathcal{B}_2&&&V = \mathcal{B}_1\oplus \mathcal{B}_2
    \end{align}
    \end{subequations}
    \subitem \textit{Sufficient conditions:}  These are also sufficient. 

    Applying the balancing condition and dualizing, we find a minimal generating set of the (1,6) cone:
    \begin{align}
        (0,0,0,0,0,0,1,-1) &\to -S(A|BCV) \nonumber\\
        (0,0,0,0,0,1,0,-1) &\to -S(B|ACV) \nonumber\\
        (0,0,0,0,1,0,0,-1) &\to -S(C|ABV) \nonumber\\
        (-1,0,0,0,0,0,1,0) &\to S(BC|V) \nonumber\\
        (-1,0,1,0,0,0,0,0) &\to S(B|V) \nonumber\\
        (-1,0,0,1,0,0,0,0) &\to S(C|V) \nonumber\\
        (0,1,0,0,0,0,-1,0) &\to S(AV) - S(BCV) \nonumber\\
        (0,-1,0,0,0,0,1,0) &\to S(BCV) - S(AV) \nonumber
    \end{align}
    \item $(1,0)$ decoupling:
    \begin{multline}\hspace*{-30pt}
        \Delta^{V,(1,0)} = \alpha_{ABV}I(A_1B_1:B_2|A_2V) + \alpha_{ACV}I(A_1C_1:C_2|A_2V) + \alpha_{ABCV}I(A_1B_1C_1:B_2C_2|A_2V)\\
        +\alpha_{BV}[S(B_1A_2V) + S(B_2V) - S(B_1B_2V) - S(A_2V)]\\
        +\alpha_{CV}[S(C_1A_2V) + S(C_2V) - S(C_1C_2V) - S(A_2V)]\\
        +\alpha_{BCV}[S(B_1C_1A_2V) + S(B_2C_2V) - S(B_1C_1B_2C_2V) - S(A_2V)]\\
        \nonumber
    \end{multline}
    \subitem \textit{Necessary conditions:} The conditions
    \begin{subequations}
    \begin{align}
    -\alpha_{ABCV} - \alpha_{ABV} - \alpha_{ACV} - \alpha_{BCV} - \alpha_{BV} - \alpha_{CV}&\geq 0 \label{5-1}\\
    -\alpha_{ABCV} - \alpha_{ACV} - \alpha_{BCV} - \alpha_{CV}&\geq 0 \label{5-2}\\
    -\alpha_{ABCV} - \alpha_{ABV} - \alpha_{BCV} - \alpha_{BV}&\geq 0 \label{5-3}\\
    -\alpha_{ABCV} - \alpha_{BCV}&\geq 0 \label{5-4}\\
    \alpha_{BCV}&\geq 0 \label{5-5}\\
    \alpha_{BV}&\geq 0 \label{5-6}\\
    \alpha_{CV}&\geq 0 \label{5-7}
    \end{align}
    \end{subequations}
    are necessary for non-positivity of $\Delta^{V,(1,0)}$. These inequalities are implied by the following states:
    \begin{subequations}
    \begin{align}
    &B_1 = C_1 = \mathcal{B}_1&&B_2 = C_2 = \mathcal{B}_2&&&V = \mathcal{B}_1\oplus\mathcal{B}_2
    \\&C_1 = \mathcal{B}_1&&C_2 = \mathcal{B}_2&&&V = \mathcal{B}_1\oplus\mathcal{B}_2
    \\&B_1 = \mathcal{B}_1&&B_2 = \mathcal{B}_2&&&V = \mathcal{B}_1\oplus\mathcal{B}_2
    \\&C_1 = \mathcal{B}_1&&B_2 = \mathcal{B}_2&&&V = \mathcal{B}_1\oplus\mathcal{B}_2
    \\&B_1 = \mathcal{B}_1, C_1 = \mathcal{B}_2&&A_2 = \mathcal{B}_3&&&V = \mathcal{B}_1\oplus \mathcal{B}_2\oplus \mathcal{B}_3
    \\&B_1 = \mathcal{B}_1&&A_2 = C_2 = \mathcal{B}_2&&&V = \mathcal{B}_1\oplus\mathcal{B}_2
    \\&C_1 = \mathcal{B}_1&&A_2 = B_2 = \mathcal{B}_2&&&V = \mathcal{B}_1\oplus\mathcal{B}_2
    \end{align}
    \end{subequations}
    \subitem \textit{Sufficient conditions:}  These are also sufficient.

    Applying the balancing condition and dualizing, we find a minimal generating set of the (1,0) cone:
    \begin{align}
        (0,0,0,0,0,0,1,-1) &\to -S(A|BCV) \nonumber\\
        (0,0,0,0,0,1,0,-1) &\to -S(B|ACV) \nonumber\\
        (0,0,0,0,1,0,0,-1) &\to -S(C|ABV) \nonumber\\
        (1,0,0,0,0,-1,0,0) &\to -S(AC|V) \nonumber\\
        (1,0,0,0,-1,0,0,0) &\to -S(AB|V) \nonumber\\
        (0,0,0,1,0,-1,0,0) &\to -S(A|CV) \nonumber\\
        (0,0,1,0,-1,0,0,0) &\to -S(A|BV) \nonumber\\
        (1,-1,0,0,0,0,0,0) &\to -S(A|V) \nonumber\\
        (-1,1,0,0,0,0,0,0) &\to S(A|V) \nonumber
    \end{align}
    \item $(4,0)$ decoupling:
    \begin{multline}
        \Delta^{V,(4,0)} = -\alpha_{AV}I(A_1:B_2|A_2V) - \alpha_{BV}I(B_1:A_2|B_2V) + \alpha_{ABCV}I(A_1B_1C_1:C_2|A_2B_2V)\\
        +\alpha_{CV}[S(C_1A_2B_2V) + S(C_2V) - S(C_1C_2V) - S(A_2B_2V)]\\
        +\alpha_{ACV}[S(A_1C_1A_2B_2V) + S(A_2C_2V) - S(A_1C_1A_2C_2V) - S(A_2B_2V)]\\
        +\alpha_{BCV}[S(B_1C_1A_2B_2V) + S(B_2C_2V) - S(B_1C_1B_2C_2V) - S(A_2B_2V)]\\
        \nonumber
    \end{multline}
    \subitem \textit{Necessary conditions:} The conditions
    \begin{subequations}
    \begin{align}
    -\alpha_{ABCV} - \alpha_{ACV} - \alpha_{BCV} - \alpha_{CV}&\geq 0 \label{6-1}\\
    \alpha_{ACV}&\geq 0 \label{6-2}\\
    \alpha_{BCV}&\geq 0 \label{6-3}\\
    \alpha_{AV}&\geq 0 \label{6-4}\\
    \alpha_{BV}&\geq 0 \label{6-5}\\
    \alpha_{CV}&\geq 0 \label{6-6}
    \end{align}
    \end{subequations}
    are necessary for non-positivity of $\Delta^{V,(4,0)}$. These inequalities are implied by the following states:
    \begin{subequations}
    \begin{align}
    &C_1 = \mathcal{B}_1&&C_2 = \mathcal{B}_2&&&V = \mathcal{B}_1\oplus\mathcal{B}_2
    \\&A_1 = \mathcal{B}_1, C_1 = \mathcal{B}_2&&B_2 = \mathcal{B}_3&&&V = \mathcal{B}_1\oplus \mathcal{B}_2\oplus \mathcal{B}_3
    \\&B_1 = \mathcal{B}_1, C_1 = \mathcal{B}_2&&A_2 = \mathcal{B}_3&&&V = \mathcal{B}_1\oplus \mathcal{B}_2\oplus \mathcal{B}_3
    \\&A_1 = \mathcal{B}_1&&B_2 = C_2 = \mathcal{B}_2&&&V = \mathcal{B}_1\oplus\mathcal{B}_2
    \\&B_1 = \mathcal{B}_1&&A_2 = C_2 = \mathcal{B}_2&&&V = \mathcal{B}_1\oplus\mathcal{B}_2
    \\&C_1 = \mathcal{B}_1&&A_2 = B_2 = \mathcal{B}_2&&&V = \mathcal{B}_1\oplus\mathcal{B}_2
    \end{align}
    \end{subequations}
    \subitem \textit{Sufficient conditions:}  These are also sufficient.

    Applying the balancing condition and dualizing, we find a minimal generating set of the (4,0) cone:
    \begin{align}
        (1,0,0,0,0,0,0,-1) &\to -S(ABC|V) \nonumber\\
        (0,0,0,1,0,0,0,-1) &\to -S(AB|CV) \nonumber\\
        (0,0,0,0,0,0,1,-1) &\to -S(A|BCV) \nonumber\\
        (0,0,0,0,0,1,0,-1) &\to -S(B|ACV) \nonumber\\
        (-1,1,0,0,0,0,0,0) &\to S(A|V) \nonumber\\
        (-1,0,1,0,0,0,0,0) &\to S(B|V) \nonumber\\
        (-1,0,0,0,1,0,0,0) &\to S(AB|V) \nonumber\\
        (1,0,0,0,-1,0,0,0) &\to -S(AB|V) \nonumber
    \end{align}
    \item $(1,2)$ decoupling:
    \begin{multline}
        \Delta^{V,(1,2)} = \alpha_VI(B_1:A_2|V) + \alpha_{ABV}I(A_1:B_2|B_1A_2V) + \alpha_{ABCV}I(A_1C_1:B_2C_2|B_1A_2V)\\
        +\alpha_{CV}[S(C_1A_2V) + S(B_1C_2V) - S(C_1C_2V) - S(B_1A_2V)]\\
        +\alpha_{ACV}[S(A_1C_1A_2V) + S(B_1A_2C_2V) - S(A_1C_1A_2C_2V) - S(B_1A_2V)]\\
        +\alpha_{BCV}[S(B_1C_1A_2V) + S(B_1B_2C_2V) - S(B_1C_1B_2C_2V) - S(B_1A_2V)]\\
        \nonumber
        \end{multline}
        
    \subitem \textit{Necessary conditions:} The conditions
    \begin{subequations}
    \begin{align}
    -\alpha_{ABCV} - \alpha_{ABV} - \alpha_{ACV} - \alpha_{BCV} - \alpha_{CV}&\geq 0\\
    -\alpha_{ABCV} - \alpha_{ACV} - \alpha_{BCV} - \alpha_{CV}&\geq 0\\
    -\alpha_{CV} - \alpha_{V}&\geq 0\\
    \alpha_{ACV}&\geq 0\\
    \alpha_{BCV}&\geq 0\\
    \alpha_{CV}&\geq 0
    \end{align}
    \end{subequations}
    are necessary for non-positivity of $\Delta^{V,(1,2)}$. These inequalities are implied by the following states:
    \begin{subequations}
    \begin{align}
    &A_1 = C_1 = \mathcal{B}_1&&B_2 = C_2 = \mathcal{B}_2&&&V = \mathcal{B}_1\oplus\mathcal{B}_2
    \\&C_1 = \mathcal{B}_1&&C_2 = \mathcal{B}_2&&&V = \mathcal{B}_1\oplus \mathcal{B}_2
    \\&B_1 = \mathcal{B}_1&&A_2 = \mathcal{B}_2&&&V = \mathcal{B}_1\oplus \mathcal{B}_2
    \\&B_1 = \mathcal{B}_1&&A_2 = \mathcal{B}_2, C_2 = \mathcal{B}_3&&&V = \mathcal{B}_1\oplus\mathcal{B}_2\oplus\mathcal{B}_3
    \\&B_1 = \mathcal{B}_1, C_1 = \mathcal{B}_2&&A_2 = \mathcal{B}_3&&&V = \mathcal{B}_1\oplus\mathcal{B}_2\oplus\mathcal{B}_3
    \\&C_1 = \mathcal{B}_1&&A_2 = B_2 = \mathcal{B}_2&&&V = \mathcal{B}_1\oplus\mathcal{B}_2
    \end{align}
    \end{subequations}
    \subitem \textit{Sufficient conditions:}  These conditions are not sufficient for $\Delta^{V,(1,2)}\leq 0$, since there are vectors in the dual of the cone that they span for which this inequality is false. That cone, whose minimal generating set is shown below, is an outer bound for the $(1,2)$ cone. Removing the generator for which $\Delta^{V,(1,2)}\leq 0$ is false (the difference of a conditional mutual information and conditional entropy at the top of the list) gives an inner bound for the $(1,2)$ cone.
    \begin{align}
        (-1,1,0,1,0,0,0,-1) &\to I(A:C|V) - S(B|ACV) \nonumber\\
        (-1,1,0,0,0,0,0,0) &\to S(A|V) \nonumber\\
        (0,1,0,0,-1,0,0,0) &\to -S(B|AV) \nonumber\\
        (0,0,0,0,0,0,1,-1) &\to -S(A|BCV) \nonumber\\
        (0,0,0,0,0,1,0,-1) &\to -S(B|ACV) \nonumber\\
        (0,0,0,0,1,0,0,-1) &\to -S(C|ABV) \nonumber\\
        (0,1,-1,0,0,0,0,0) &\to S(AV) - S(BV) \nonumber\\
        (0,-1,1,0,0,0,0,0) &\to S(BV) - S(AV) \nonumber
    \end{align}
    
    \item $(1,4)$ decoupling:
    \begin{multline}
        \Delta^{V,(1,4)} = \alpha_VI(A_1B_1:A_2|V) + \alpha_{ABCV}I(C_1:B_2C_2|A_1B_1A_2V)\\
        +\alpha_{BV}[S(B_1A_2V) + S(A_1B_1B_2V) - S(B_1B_2V) - S(A_1B_1A_2V)]\\
        +\alpha_{CV}[S(C_1A_2V) + S(A_1B_1C_2V) - S(C_1C_2V) - S(A_1B_1A_2V)]\\
        +\alpha_{ACV}[S(A_1C_1A_2V) + S(A_1B_1A_2C_2V) - S(A_1C_1A_2C_2V) - S(A_1B_1A_2V)]\\
        +\alpha_{BCV}[S(B_1C_1A_2V) + S(A_1B_1B_2C_2V) - S(B_1C_1B_2C_2V) - S(A_1B_1A_2V)]\\
        \nonumber
    \end{multline}
    
    \subitem \textit{Necessary conditions:} The conditions
    \begin{subequations}
    \begin{align}
    -\alpha_{ABCV} - \alpha_{ACV} - \alpha_{BCV} - \alpha_{CV}&\geq 0\\
    -\alpha_{BCV} - \alpha_{BV} - \alpha_{CV} - \alpha_{V}&\geq 0\\
    \alpha_{ACV}&\geq 0\\
    \alpha_{BCV} &\geq 0\\
    \alpha_{BV}&\geq 0\\
    \alpha_{CV}&\geq 0
    \end{align}
    \end{subequations}
    are necessary for non-positivity of $\Delta^{V,(1,4)}$. These inequalities are implied by the following states:
    \begin{subequations}
    \begin{align}
    &C_1 = \mathcal{B}_1&&C_2 = \mathcal{B}_2&&&V = \mathcal{B}_1\oplus\mathcal{B}_2
    \\&A_1 = \mathcal{B}_1&&A_2 = \mathcal{B}_2&&&V = \mathcal{B}_1\oplus\mathcal{B}_2
    \\&B_1 = \mathcal{B}_1&&A_2 = \mathcal{B}_2, C_2 = \mathcal{B}_3&&&V = \mathcal{B}_1\oplus\mathcal{B}_2\oplus\mathcal{B}_3
    \\&B_1 = \mathcal{B}_1, C_1 = \mathcal{B}_2&&A_2 = \mathcal{B}_3&&&V = \mathcal{B}_1\oplus\mathcal{B}_2\oplus\mathcal{B}_3
    \\&A_1 = C_1 = \mathcal{B}_1&&B_2 = \mathcal{B}_2&&&V = \mathcal{B}_1\oplus \mathcal{B}_2
    \\&C_1 = \mathcal{B}_1&&A_2 = B_2 = \mathcal{B}_2&&&V = \mathcal{B}_1\oplus\mathcal{B}_2
    \end{align}
    \end{subequations}
    \subitem \textit{Sufficient conditions:}  These conditions are not sufficient for $\Delta^{V,(1,4)}\leq 0$, since there are vectors in the dual of the cone that they span for which this inequality is false. That cone, whose minimal generating set is shown below, is an outer bound for the $(1,4)$ cone. Removing the generator for which $\Delta^{V,(1,4)}\leq 0$ is false (the difference of a conditional mutual information and conditional entropy at the top of the list) gives an inner bound for the $(1,4)$ cone.
    \begin{align}
        (-1,1,0,1,0,0,0,-1) &\to I(A:C|V) - S(B|ACV) \nonumber\\
        (-1,1,0,0,0,0,1,-1) &\to I(A:BC|V) \nonumber\\
        (-1,1,0,0,0,0,0,0) &\to S(A|V) \nonumber\\
        (0,1,0,0,-1,0,0,0) &\to -S(B|AV) \nonumber\\
        (0,-1,0,0,1,0,0,0) &\to S(B|AV) \nonumber\\
        (0,0,0,0,0,1,0,-1) &\to -S(B|ACV) \nonumber\\
        (0,0,0,0,0,0,1,-1) &\to -S(A|BCV) \nonumber\\
        (0,1,0,0,0,0,0,-1) &\to -S(BC|AV) \nonumber
    \end{align}
    
\end{enumerate}

\end{document}